\documentclass[aps,pre,twocolumn,groupedaddress,floatfix,longbibliography,superscriptaddress]{revtex4-1}
\usepackage{amsmath}
\usepackage{amssymb}
\usepackage{amsfonts}
\usepackage{graphicx}
\usepackage{bbold}
\usepackage{bm}
\usepackage{bm}
\usepackage{cancel}
\usepackage{times,float}
\usepackage{graphicx}
\usepackage[usenames,dvipsnames,svgnames]{xcolor}
\usepackage{hyperref}
\hypersetup{colorlinks=true, linkcolor=Blue, citecolor=PineGreen,urlcolor=cyan}
\usepackage{multirow}
\usepackage{ulem}
\usepackage{braket}

\newcommand{\bea}{\begin{eqnarray}}
\newcommand{\eea}{\end{eqnarray}}
\newcommand{\nn}{\nonumber}

\begin{document}
\title{Ground state and polarization of an hydrogen-like atom near a Weyl semimetal}
\author{Jorge David Casta\~no-Yepes}
\email{jcastano@fis.uc.cl}
\affiliation{Instituto de Física, Pontificia Universidad Católica de Chile, Vicuña Mackenna 4860, Santiago, Chile.}
\author{D. J. Nader}
\email{daniel\_nader@brown.edu}
\address{Facultad de F\'isica, Universidad Veracruzana, Apartado Postal 91-090, Xalapa, Veracruz, M\'exico}
\affiliation{Department of Chemistry, Brown University, Providence, Rhode Island 02912, United States}
\author{A. Mart\'in-Ruiz}
\email{alberto.martin@nucleares.unam.mx}
\address{Instituto de Ciencias Nucleares, Universidad Nacional Aut\'{o}noma de M\'{e}xico, 04510 Ciudad de M\'{e}xico, M\'{e}xico}


\begin{abstract}
In this paper we study the effects of a topological Weyl semimetal (WSM) upon the ground state and polarization of an hydrogen-like atom near its surface. The WSM is assumed to be in the equilibrium state and at the neutrality point, such that the interaction between the atomic charges and the material is fully described (in the non retarded regime) by axion electrodynamics, which is an experimentally observable signature of the anomalous Hall effect in the bulk of the WSM. The atom-WSM interaction provides additional contributions to the Casimir-Polder potential thus modifying the energy spectra and wave function, which now became distance dependent. Using variational methods, we solve the corresponding Schr\"{o}dinger equation for the atomic electron. The ground state and the polarization are analyzed as a function of the atom-surface distance, and we directly observe the effects of the nontrivial topology of the material by comparing our results with that of a topologically trivial sample. We also study the impact of the medium's permittivity by assuming a hydrogen atom in vacuum, and a donor impurity in the semiconductors gallium arsenide (GaAs) and gallium phosphide (GaP). We found that the topological interaction behaves as an effective-attractive charge so that the electronic cloud tends to be polarized to the interface of materials. Moreover, the loss of wave-function normalization is interpreted as a critical location from below which the bound state is broken.
\end{abstract}
\pacs{xxx xxx xx xx}
\maketitle
\section{Introduction}

Topological materials have attracted great attention recently both from the theoretical and experimental sides. Topological insulators (TIs) are characterized by a gapped bulk and gapless boundary states that are robust against disorder \cite{RevModPhys.83.1057}. Further, Weyl semimetals (WSMs) are phases with broken time-reversal or spatial-inversion symmetry, whose electronic structure contains pairs of band crossing points (Weyl nodes) in the Brillouin zone provided the Fermi level is close to the Weyl nodes \cite{RevModPhys.90.015001}. Besides their spectroscopic distinguishing features, these phases also exhibit unusual electromagnetic responses, which are described by topological field theories \cite{PhysRevB.78.195424,PhysRevB.86.115133} akin to axion electrodynamics \cite{PhysRevLett.58.1799}.

On the other hand, the manipulation of charge carriers in materials by dopping has become of central interest in the technologies that enable the semiconductor electrical conductivity control over several orders of magnitude~\cite{sun2014significant,van2010controlling,o1998electron, bellotti1999ensemble,chern2006excitation}. The conductivity enhancement is understood in terms of bound states associated with the impurity that, after on an excitation, become delocalized as conduction or valence band states. Although the Coulomb potential of donor/acceptor may scatter the mobile charges and therefore reduces its mobility, the combination of large effective dielectric constant and small effective masses in a semiconductor medium result in wave functions extended over a large space, which implies binding energies of a few electronvolts~\cite{kohn1957shallow}. In bulk semiconductors such as Si, GaAs or GaP, the potential of charged impurities is screened by the dielectric response of the environment, generating localized Bloch states with hydrogen-like wave functions~\cite{xie2017demonstration, PhysRevLett.110.166404,PhysRevB.68.233204,bassani1974electronic}.


It is well-established that a hydrogenic donor/acceptor impurity is entirely equivalent to a hydrogen atom regarding their quantum mechanically description at the effective-mass approximation~\cite{RevModPhys.50.797,kohn1957solid}. From the theoretical point of view, the donor levels into a semiconductor material have been attractive in order to explore the optical properties and modifications to the band structure.  Bastard performed a variational calculation of the binding energy for hydrogenic impurity states in a quantum well, where the energy levels depend on the position and the well's thickness~\cite{PhysRevB.24.4714}. Keldysh showed that the Coulomb interaction is sensitive to the impurity's location in systems with interfaces between two materials~\cite{Keldysh1979}. Lipari solved the effective mass equations for a donor impurity due to its interface's distance of the semiconductor-insulator juncture~\cite{lipari1978electronic}. Moreover, the experimental developments that make possible the introduction and manipulation of impurities in low-dimensional systems opened the study of their effects in the so-called quantum dots~\cite{PhysRevB.58.R15997,yu1996optical,PhysRevB.50.7602,PhysRevLett.72.416,bhargava1994doped,gallagher1995homogeneous}. Such experimental setups have inspired theoretical search of the hydrogenic-like impurities in confined nanosystems: Banin {\it et al.} developed a method to dope semiconductor nanocrystals with metallic impurities finding that a low concentration of donor impurities the red-shift on the photoluminescence spectrum is well explained when considering both donor and acceptor hydrogenic impurities~\cite{mocatta2011heavily}. Baimuratov {\it et al.} discussed the level anticrossing for impurity donor states in a spherical semiconductor nanocrystal~\cite{baimuratov2014level}. Mughnetsyan {\it et al.} calculated electric and magnetic fields' effect on the binding energy and photoionization cross-section on an of--axis hydrogen donor impurity located in a quantum well-wire~\cite{mughnetsyan2008binding}, and recently, Aghajanian {\it et al.} observed localized states described by hydrogen wave functions in the valence band's edge for dopped two-dimensional semiconductors~\cite{PhysRevB.101.081201}. The research on that kind of impurities also covers areas such as quantum information, where the spin-orbit interaction plays a crucial role in the impurities' hyperfine structure.~\cite{PhysRevLett.88.027903,kane2000silicon,rosero2019spin}.

In the aforementioned works, the juncture is always between materials with trivial band structures, i.e., they have no topological features. Then, it results interesting to study situations when a topologically ordered material interacts with a trivial one, and search for signatures of the topological nontriviality into the physical observables. Some works have been done in this spirit. For example, some particular classical electrodynamics configurations both in TIs~\cite{PhysRevA.92.063831, PhysRevD.92.125015, PhysRevD.93.045022, PhysRevD.94.085019, PhysRevD.98.056012, PhysRevD.99.116020,Medel_23} and WSMs \cite{Kargarian2015, PhysRevLett.117.217204, PhysRevB.99.155142, PhysRevB.93.241402, PhysRevB.92.201407}, the frequency shift induced by the Casimir-Polder interaction between atoms and TIs~\cite{PhysRevA.95.023805,  PhysRevA.97.022502, Bonilla}, and the optical absorption of semiconductor-TI quantum dots~\cite{CASTANOYEPES2020114202, PhysRevA.102.013720, Castro_Enriquez_2020, SciRep22}, among others. Moved by these researches, in this paper we investigate the effects of a Weyl semimetallic phase upon hydrogenlike ions near its surface, taking into account the modifications arising from the topological nontriviality of the material. Taking a WSM in the equilibrium state and at the neutrality point, we avoid undesired contributions such as the chiral magnetic effect and the chiral separation effect, thus allowing us to concentrate on the consequences of the bulk anomalous Hall effect upon the ground state and polarization of a hydrogen-like atom located near its surface. In the nonretarded regime, our model Hamiltonian includes the electromagnetic interaction between the Hall currents in the bulk of the WSM and the atomic electron. By means of variational methods we solve the Schr\"{o}dinger's equation for atomic electron and study, as a function of the atom-surface distance, the ground state and the atomic polarization.

The paper is organized as follows. In Sec.~\ref{sec:Electromagnetic_response_of_WSM} we review the basics of the electromagnetic response of WSM. The Hamiltonian describing the interaction between the hydrogen-like atom and the WSM is derived in Sec.~\ref{sec:Hydrogenlike_ion_near_the_WSM_surface}. Sec.~\ref{sec:Variational_Functions} presents the form of the variational functions we use, and Sec.~\ref{sec:Results_and_Discussion} presents the corresponding results and discussion. Finally, the summary and conclusions can be found in Sec.~\ref{sec:summary_and_Conclusions}.

\section{Electromagnetic response of WSM}\label{sec:Electromagnetic_response_of_WSM}

The low-energy effective field theory governing the electromagnetic response of WSMs with a single pair of band-touching points, independently of the microscopic details, is defined by the usual Maxwell Lagrangian density \cite{PhysRevB.86.115133}
\begin{align}
    \mathcal{L} _{\mbox{\scriptsize Max}} = \frac{1}{2} \left[ \epsilon {\bf{E}} ^{2} - (1/ \mu ) {\bf{B}} ^{2} \right] - \rho  \phi + {\bf{A}} \cdot {\bf{J}} ,  \label{MaxwellL}
\end{align}
supplemented by an additional $\theta$-term of the form
\begin{align}
    \mathcal{L} _{\theta} = \frac{\alpha}{4 \pi ^{2}} \theta ({\bf{r}},t) \, {\bf{E}} \cdot {\bf{B}} . \label{ThetaL}
\end{align}
The so-called axion field $\theta ({\bf{r}},t)$ has the following form
\begin{align}
    \theta ({\bf{r}},t) = 2 {\bf{b}} \cdot {\bf{r}} - 2 b _{0} t , \label{Theta}
\end{align}
where $2 {\bf{b}}$ is the separation between the Weyl nodes in momentum space, $2 b _{0}$ is their energy offset, and $\alpha \simeq 1/ 137$ is the fine structure constant.  It should be noted that unlike to topological insulators for which $\theta$ is quantized due to time-reversal symmetry \cite{PhysRevB.78.195424}, in WSMs the nonquantized expression for $\theta$ is due to the time-reversal symmetry breaking by ${\bf{b}}$ and the inversion symmetry breaking by $b _{0}$ \cite{PhysRevB.86.115133}.

The physical manifestations of the $\theta$ term can be best understood from the associated field equations, which give rise to the following charge density and current density response,
\begin{align}
    \rho _{\theta} ({\bf{r}},t) &= \frac{\delta \mathcal{L} _{\theta}}{\delta \phi} = - \frac{\alpha}{2 \pi ^{2}} {\bf{b}} \cdot {\bf{B}} , \label{ChargeDensity} \\ {\bf{J}} _{\theta} ({\bf{r}},t) &= \frac{\delta \mathcal{L} _{\theta}}{\delta {\bf{A}}} = \frac{\alpha }{2 \pi ^{2}} \left(  {\bf{b}} \times {\bf{E}} - b _{0} {\bf{B}} \right) . \label{CurrentDensity}
\end{align}
The charge density of Eq.~(\ref{ChargeDensity}) together with the first term of the current density in Eq.~(\ref{CurrentDensity}) encode the anomalous Hall effect, which is expected to occur in a WSM with broken TR symmetry \cite{PhysRevB.84.075129, PhysRevLett.107.127205}. The chiral magnetic effect, which manifests in WSMs with broken I symmetry, indicates that a ground state dissipationless current is generated along a static magnetic field even in the absence of electric fields \cite{Landsteiner}. One part of this peculiar phenomenon is described by the $b_{0}$-dependent term in the current density given by Eq.~(\ref{CurrentDensity}).

The electromagnetic response of WSMs is not fully captured by axion electrodynamics, but as in ordinary metals, there are additional currents which depend linearly on the electric and magnetic fields. For example, being a WSM a metallic system, the Ohm's law still holds. If we have chiral fermions with chemical potentials $\mu _{L}$ and $\mu _{R}$ for left- and right-handed fermions, driven by a single frequency electric field, there is a term of the form $J _{i} = \sigma _{ij} (\omega ) E _{j}$, where $\sigma _{ij} (\omega )$ is the longitudinal conductivity tensor given by 
\begin{align}
\sigma _{ij} (\omega ) = \frac{e ^{2} \tau}{6 \pi ^{2} \hbar ^{3} v _{F}}  \frac{\delta _{ij}}{1 + \mathrm{i} \omega \tau } ( \Lambda _{L} ^{2} + \Lambda _{R} ^{2} ) , \label{LongConductivity}
\end{align}
where $e$, $v_{F}$ and $\tau$ are the electron charge, Fermi velocity and scattering time,  respectively. $\Lambda _{\chi} \equiv \mu _{\chi} - b _{0 \chi}$ is the filling of the cone with chirality $\chi$. In the Appendix \ref{AppKinetic} we derive the formula (\ref{LongConductivity}) by using kinetic theory. The corresponding zero temperature carrier density is found to be
\bea
n = \frac{ \Lambda _{L} ^{2} + \Lambda _{R} ^{2}}{2 \pi ^{2} (\hbar v _{F}) ^{3}}.
\eea

Clearly, $\sigma _{ij}$ varies as the square of the filling, and therefore it vanishes exactly at the Weyl nodes. This is expected since the density of states in this model vanishes when approaching the Weyl point.

In addition, there are two additional current terms depending on the magnetic field, namely,
\begin{align}
    {\bf{J}} = \frac{\alpha }{2 \pi ^{2}} \mu _{5} {\bf{B}} , \qquad {\bf{J}} _{5} = \frac{\alpha }{2 \pi ^{2}} \mu {\bf{B}} , \label{CurrentMagneticF}
\end{align}
where $\mu _{5} = (\mu _{L} - \mu _{R})/2$ and $\mu = (\mu _{L} + \mu _{R})/2$ are the chiral and electric chemical potentials, respectively. The second part of the chiral magnetic effect is given by ${\bf{J}} $ in Eq. (\ref{CurrentMagneticF}), which arises from an imbalance between chemical potentials of right- and left-handed fermions. The total contribution to the CME current is then \cite{Landsteiner}
 \begin{align}
    {\bf{J}} _{\mbox{\scriptsize CME}} = \frac{\alpha }{2 \pi ^{2}} ( \mu _{5} - b _{0}) {\bf{B}} = \frac{\alpha }{4 \pi ^{2}} (\Lambda _{L} + \Lambda _{R}) {\bf{B}}, \label{CME-Current}
\end{align}
that vanishes for $b _{0} = \mu _{5} $ in which case the WSM is said to be at the equilibrium state. On the other hand, ${\bf{J}} _{5}$ in Eq. (\ref{CurrentMagneticF}) that is identified with the chiral separation effect, vanishes for $\mu = 0$, condition that defines the neutrality point. The existence of the static chiral magnetic effect is, however, ruled out in crystalline solids \cite{PhysRevLett.111.027201}, which is also consistent with our understanding that static
magnetic fields do not generate equilibrium currents. All in all, the full electromagnetic response of a WSM is described by axion electrodynamics defined by the Lagrangian $\mathcal{L} _{\mbox{\scriptsize Max}} + \mathcal{L} _{\theta}$, together with the afore discussed current terms. In summary, in the presence of electric and magnetic fields, the Weyl response is described by the anomalous Hall effect, the Ohm's law, the chiral magnetic effect and the chiral separation effect. As we shall discuss below, in the problem at hand the nontopological contributions can overwhelm the topological ones, and hence, we need to choose appropriately the setup in order to obtain physical effects due to the topologically nontrivial Weyl nodes.

\section{Hydrogen-like ion near the WSM surface}\label{sec:Hydrogenlike_ion_near_the_WSM_surface}

\subsection{Statement of the problem}

In general, the interaction between a WSM and an atom nearby will be dominated by the trivial optical properties of the material, such as the longitudinal conductance. To be precise, the anomalous Hall current in Eq.~(\ref{CurrentDensity}) will be overwhelmed by the nontopological Ohm's current, and so would its contribution to the interaction between the atom and the WSM. In order to avoid this problem and disentangling the topological from the trivial contributions, we have to make some simplifying but realistic assumptions. On the one hand, it is well known that the static chiral magnetic effect is ruled out in crystalline solids \cite{PhysRevLett.111.027201}, although it can be realized under nonequilibrium circumstances. Therefore, being this a static problem, we can safely take a WSM in the equilibrium state, i.e. with $b _{0} = \mu _{5} $. On the other hand, as suggested by Eq. (\ref{LongConductivity}), the longitudinal conductivity goes out when the filling of the cones equals to zero, i.e. for $\Lambda _{L} = \Lambda _{R} = 0 $. This is reasonable, since in WSMs, the carrier density $n$ is typically very low since the Fermi momentum is small around the Weyl nodes. When this happens, the Ohmic conductivity can be ignored, and we can set $\sigma _{ii} = 0$. If $n$ is increased the conductivity can no longer be ignored and the Ohm's current cannot be set to zero. So, for definiteness, we take the fillings equal to zero. Clearly, this condition simultaneously guarantees the vanishing of the chiral magnetic effect.

\subsection{Model Hamiltonian} \label{ModelHamiltonian}

The interaction between atoms and surfaces have proven to be of fundamental importance in physics. For example, atom-surface interactions play an important role in atomic force microscopy and they also affect the properties of an atom or molecule nearby. { In the particular case of metallic surfaces and/or dielectric samples, since the interaction with an atom takes place far from the surface (as compared with the atom size), the atom-surface interaction can be modeled by nonretarded electrostatic forces and the matter can be treated as continuum with a well-defined frequency-dependent dielectric function. Under this circumstance, the only force relevant in the problem is the Casimir-Polder force acting upon the electron, thus affecting its quantum properties, such as the energy levels for a given quantum state and the decay rates of excited states, which now become functions of the atom-surface distance. In the nonretarded regime, the atom-surface interaction can be modelled by the electrostatic method of images, i.e. the images of the electric charges of the atom act as another atom which exerts additional forces on the atomic electron \cite{Ganesan_1996, Simonovic_1997, DUNNING200369, SIMONOVIC200460}. }

If the material body is, for example, a topological insulator, additional interactions arise due to the topological magnetoelectric effect: the charges in the atom will induce, besides image electric charges, image magnetic monopoles as well (physically induced by a vortex Hall current in the surface), which in turn will interact with the atom via the minimal coupling prescription. This problem has been considered within the framework of quantum \cite{PhysRevA.97.022502, Bonilla} and classical \cite{Mart_n_Ruiz_2017} mechanics. The aim of this work is to built up a formalism that allows us to investigate the influence of a topological WSM upon an atom nearby. Due to the broken symmetries in the bulk, additional nontrivial topological effects may result as compared to the case of the TIs. So, we first recall the problem of an electric charge placed near to a WSM half-space, {such that the WSM's electromagnetic response is described by Maxwell macroscopic theory supplemented with the axionic term characteristic of the topological phases.}

\begin{figure}[h!]
    \centering
    \includegraphics[scale=0.4]{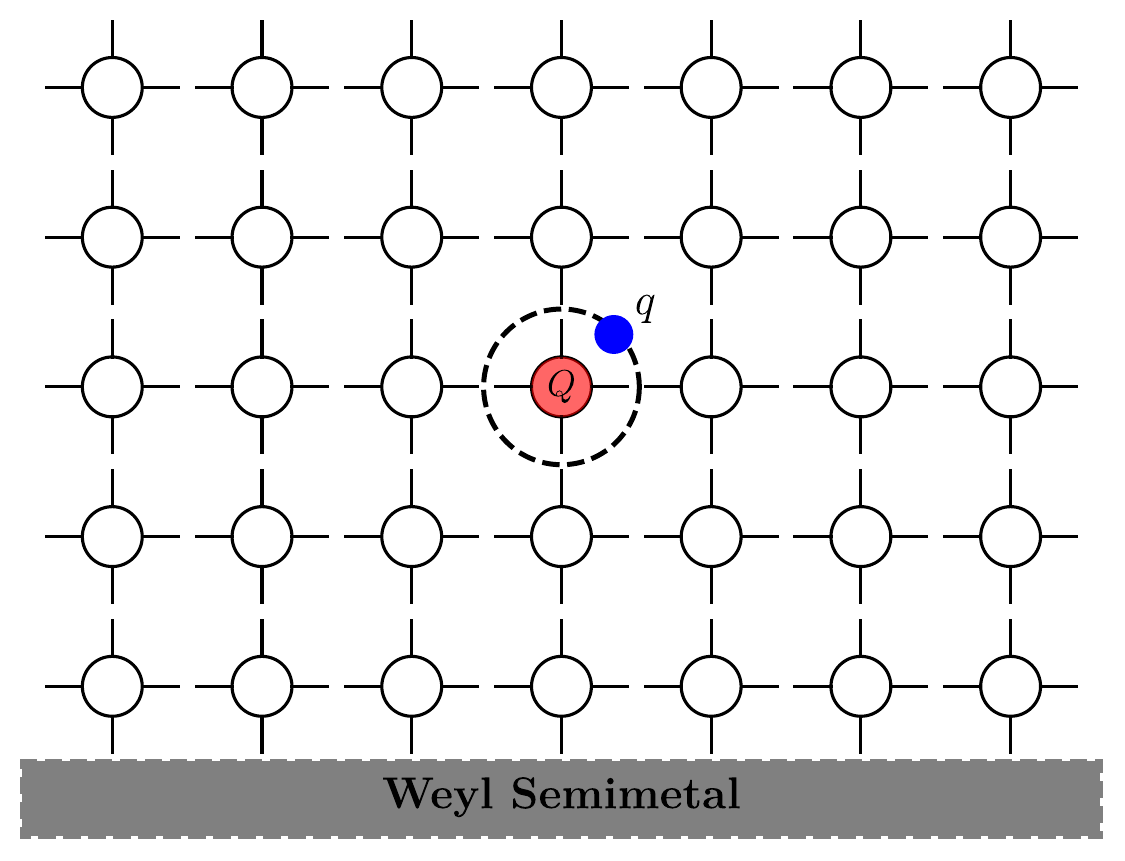}
    \caption{Hydrogenic impurity near a WSM. The hydrogen-like atom is embedded in a material of permittivity $\epsilon_2$. The charge $q$ is negative for donors and positive for acceptors and vice-versa for the charge $Q$. }
    \label{Fig:Impurity_WSM}
\end{figure}

The electromagnetic response of TIs is rather simple, since the only nontrivial physical effect is a half-quantized quantum Hall effect on the sample’s surfaces. However, in the case of WSMs, Eq.~(\ref{ThetaL}) does modify the field equations in the bulk and thus provides additional observable consequences, namely, the anomalous Hall effect. Consider the geometry presented in  Fig.~\ref{Fig:Impurity_WSM}. The lower half-space ($z<0$) is occupied by a topological WSM with a pair of nodes separated along the $k_{z}$-direction in the bulk Brillouin zone, while the upper half-space ($z>0$) is occupied by a dielectric/semiconductor. { Being this a static problem, we neglect all frequency dependences to the conductivities and permittivities, such that the lower half-space is just a material that is solely a bulk Hall material with current response given by the Hall conductivity $\sigma _{xy}$ and the dielectric constant $\epsilon _{1}$, and the upper half-space is characterized solely by its dielectric constant $\epsilon _{2}$.} An electric charge of strength $q$ is brought at a distance $z _{0}>0$ from the surface $z=0$. Working in cylindrical coordinates $(\rho , \varphi , z )$ to exploit the axial symmetry of the problem, in the region $z>0$ the  electric potential is found to be \cite{PhysRevB.99.155142}
\begin{align}
\Phi _{q} ( {\bf{r}} , {\bf{r}} _{0} ) &= \frac{q}{\epsilon _{2}} \frac{1}{\vert {\bf{r}} - {\bf{r}} _{0} \vert } + \frac{q}{\epsilon _{2}} \frac{\epsilon _{2} - \epsilon _{1}}{\epsilon _{2} + \epsilon _{1}} \frac{1}{\vert {\bf{r}} + {\bf{r}} _{0} \vert } - \frac{2 q \epsilon _{1}}{\epsilon _{1} + \epsilon _{2}} \times \notag \\  & \phantom{=} \int _{0} ^{\infty} \!\! \frac{ ( \alpha _{+} ^{2} + \alpha _{-} ^{2} - k ^{2} ) J _{0} (k \rho) e ^{- k ( z + z _{0} )}  }{\epsilon _{1} \! \left( \alpha _{+} ^{2} + \alpha _{-} ^{2} \right) + \epsilon _{2} k ^{2} +  k  \alpha _{+} \! \left( \epsilon _{1} \! + \! \epsilon _{2} \right)} dk  . \label{EscPot}
\end{align}
where ${\bf{r}} _{0} = z _{0} \hat{{\bf{e}}} _{z}$, $J _{n}$ is the $n$th order Bessel function of the first kind, $\rho = \sqrt{x ^{2} + y ^{2}} $, and
\begin{align}
\alpha _{\pm} ( k ) = \sqrt{\frac{k}{2} \left( \sqrt{k ^{2} + \Sigma ^{2}} \pm k \right)} , \label{alpha}
\end{align}
where $\Sigma = \frac{4 \pi}{c} \frac{\sigma _{xy}}{\sqrt{\epsilon _{1}}}$ is an effective bulk Hall conductivity (with dimensions of inverse length). Clearly, the electric potential can be interpreted as due to the original electric charge of strength $q$ at $z _{0}$, an image electric charge of strength $q (\epsilon _{2} - \epsilon _{1}) / (\epsilon _{2} + \epsilon _{1})$ at $- z _{0}$, and an additional term arising from the nontrivial topology of the WSM. In the limit $\sigma _{xy} \to 0$ the last term in Eq.~(\ref{EscPot}) vanishes, and the potential can be interpreted solely in terms of the image charge. Due to the axial symmetry, the vector potential has the form ${\bf{A}} = A _{q} ( \rho , \theta ) \hat{{\bf{e}}} _{\varphi} $, choice that naturally  naturally satisfies the Coulomb gauge. In the problem at hand, the function $A _{q} ( \rho , \theta )$ for the region $z>0$ becomes
\begin{align}
A _{q} ( {\bf{r}} , {\bf{r}} _{0} ) = \! \int _{0} ^{\infty} \! \frac{ 2 q \epsilon _{1} \alpha _{-} k J _{1} (k \rho) e ^{- k ( z + z _{0} )}   }{\epsilon _{1} \! \left( \alpha _{+} ^{2} + \alpha _{-} ^{2} \right) + \epsilon _{2} k ^{2} +  k  \alpha _{+} \! \left( \epsilon _{1} \! + \! \epsilon _{2} \right)} dk  . \label{VecPot}
\end{align}
Clearly, the corresponding magnetic field arises from the topological nontrivality of the material, i.e. a direct manifestation of the anomalous Hall effect. It vanishes in the limit $\sigma _{xy} \to 0$, when magnetoelectricity disappears. Physically, the induced magnetic field can be interpreted as generated by an infinite number of $2+1$ Dirac subsystems (one for each value of $z$ in the bulk) supporting a surface Hall current \cite{PhysRevB.99.155142}. Therefore, the magnetic field, as well as the $\Sigma$-dependent term of the electric field, cannot be interpreted in terms of a well-localized image source.

With the help of the scalar and vector potentials above, we are ready to write down the interaction Hamiltonian between a WSM and an atom nearby. {To this end, some assumptions are needed, that we shall discuss in the following. For the sake of simplicity we expressly consider the case of an atom located near to the surface with no arc states. The analogous problem of an atom located in front of a surface that supports Fermi arcs would also be of great interest. However, from a practical point of view, we assume that the WSM phase has been properly characterized, such that the surfaces with/without arc states have been identified. For example, when a WSM phase is produced from a Dirac semimetal by applying an external magnetic field, the separation between nodes will be along the field direction and thus the identification of the surfaces supporting arc states is possible. Therefore, we can safely choose the configuration depicted in Fig. \ref{Fig:Impurity_WSM}, and we left the complementary problem for future investigations. Last but not least, we have to justify the validity the dielectric response picture. The interaction between an atom and a material body (e.g. conductor, dielectric or topological insulator) depends on the distance between them. As long as the atom-body separation is sufficiently large compared with the atomic radius on the one hand, and the typical distance between the atomic constituents of the body on the other hand, the atom-body interaction can be calculated within the frame of macroscopic electrodynamics, provided there is no direct wave-function overlap. If the atom is close to the surface (at least of a few atomic radii), the interaction is dominated by electrostatics. However, retardation becomes important for atoms further away from the surface. Experimental support for the use of macroscopic electrodynamics in these systems is found in Refs. \cite{Oria_1991, PhysRevLett.68.3432, PhysRevLett.86.2766,PhysRevLett.83.5467, PhysRevLett.70.560, PhysRevLett.86.987}. }

In the nonretarded regime, the atom-surface interaction is achieved by computing the Coulomb interaction between all atomic charges and all image charges \cite{Ganesan_1996, Simonovic_1997, DUNNING200369, SIMONOVIC200460}. In the problem at hand we cannot interpret the electric field in terms of localized image charges, but we are able to calculate the electrostatic interaction energy with the help of Eq.~(\ref{EscPot}). Due to the anomalous Hall effect of the WSM, the atomic charges will also produce magnetic fields sourced by nonlocalized distributions in the bulk \cite{PhysRevB.99.155142}, which in turn will interact with the atomic electron. Therefore, in the minimal coupling prescription, the quantum Hamiltonian we shall consider reads
\begin{align}
    \hat{H} = \frac{1}{2 \mu } \left( \hat{{\bf{p}}}  - \frac{e}{c} {\bf{A}} \right) ^{2} + V ({\bf{r}}) ,  \label{Hamiltonian}
\end{align}
where $\mu$ is the mass of the moving charge, $c$ is the speed of  light and $\hat{{\bf{p}}} = - i \hbar \nabla $. In Eq.~(\ref{Hamiltonian}), $V ({\bf{r}})$ accounts for the electrostatic interactions and ${\bf{A}}$ is the vector potential. Let us derive these terms.

Treating the ion as an electric composite system, the effective charge density can be expressed as $\rho ({\bf{r}} ^{\prime}) = e [ Z \, \delta ({\bf{r}} ^{\prime} - {\bf{r}} _{0}) - \delta ({\bf{r}} ^{\prime} - {\bf{r}} - {\bf{r}} _{0}) ]$, where $Z$ is the atomic number and ${\bf{r}}$ localizes the atomic electron from the nucleus. The electric field due the ion can thus be computed by superposing the solution (\ref{EscPot}). So, the interaction energy between the hydrogen-like atom and the WSM can be written as
\begin{align}
    V ( \rho , \theta ) &= \frac{1}{2} \Big[ - e \phi _{Ze} ( {\bf{r}} + {\bf{r}} _{0} , {\bf{r}} _{0} ) - e \phi _{-e} ( {\bf{r}} + {\bf{r}} _{0} , {\bf{r}} + {\bf{r}} _{0} ) \notag \\ & \hspace{0.7cm} + Ze  \phi _{Ze} ( {\bf{r}} _{0} , {\bf{r}} _{0} ) + Ze  \phi _{-e} ( {\bf{r}} _{0} , {\bf{r}} + {\bf{r}} _{0} ) \Big]  , \label{PotEnergy}
\end{align}
where $\phi _{q} ( {\bf{r}} , {\bf{r}} ^{\prime})$ is the scalar potential at the position ${\bf{r}}$ due to a charge $q$ at ${\bf{r}} ^{\prime}$, given by Eq.~(\ref{EscPot}). Clearly, the potential energy (\ref{PotEnergy}) accounts for the many pairwise interactions in our configuration. For example, the first term, $- e \phi _{Ze} ( {\bf{r}} + {\bf{r}} _{0} , {\bf{r}} _{0} )$, corresponds to the interaction energy between the nucleus and the atomic electron, including the contributions arising from the presence of the WSM. The second term, $- e \phi _{-e} ( {\bf{r}} + {\bf{r}} _{0} , {\bf{r}} + {\bf{r}} _{0} ) $, is the electron-electron interaction energy, which includes a divergent term arising from the electron self-energy, which we discard. The third term, $Ze  \phi _{Ze} ( {\bf{r}} _{0} , {\bf{r}} _{0} )$, is the nucleus-nucleus interaction energy, which contains also a divergent terms due to the nucleus self-energy, which we discard. The last term, $Ze  \phi _{-e} ( {\bf{r}} _{0} , {\bf{r}} + {\bf{r}} _{0} )$, is the interaction energy between the electron and the nucleus (which is the same that the first term). In a coordinate system attached to the nucleus, the potential energy (\ref{PotEnergy}) takes the form: 
\begin{widetext}
\begin{align}
    V ( \rho , \theta ) &= - \frac{Z e ^{2}}{\epsilon _{2}  r}  - \frac{ e ^{2} }{\epsilon _{2}} \frac{\epsilon _{2} - \epsilon _{1}}{\epsilon _{2} + \epsilon _{1}} \left(\frac{Z}{\sqrt{r ^{2} + 4 z _{0} ( z + z _{0})}} - \frac{1}{4(z + z _{0})}\right) +  \frac{ e ^{2} \epsilon _{1}}{\epsilon _{1} + \epsilon _{2} } \mathcal{I} ( \rho , \theta ) .  \label{PotEnergy2}
\end{align}
where
\begin{align}
    \mathcal{I} ( \rho , \theta ) &= \int _{0} ^{\infty}  dk \, \frac{ ( \alpha _{+} ^{2} + \alpha _{-} ^{2} - k ^{2} ) }{\epsilon _{1} \! \left( \alpha _{+} ^{2} + \alpha _{-} ^{2} \right) + \epsilon _{2} k ^{2} +  k  \alpha _{+} \! \left( \epsilon _{1} \! + \! \epsilon _{2} \right)} \left[2 Z J_0\left(k \rho \right)e^{-k(z+2z_0)}-e^{-2k(z+z_0)} \right] . 
\end{align}
\end{widetext}
The first term in Eq.~(\ref{PotEnergy2}) is the usual Coulomb interaction experienced by the atomic electron due to the nucleus. The second term corresponds to the interaction between the image nucleus and the atomic electron, while the third term is the interaction between the electron and its own image. The last term, which cannot be interpreted in terms of images, is a direct manifestation of the anomalous Hall effect. 

The vector potential can be computed in a similar fashion. However, from the result of Eq.~(\ref{VecPot}), we observe that the vector potential vanishes along the line perpendicular to the charge source, i.e. for $\rho = 0$. This means that the vector potential sourced by the atomic electron does not act upon the electron itself. Therefore, the only vector potential to be considered is that sourced by the nucleus. So, in the coordinate system attached to the nucleus the nonzero component of the vector potential reads
\begin{align}
    A _{Ze} ( \rho , \theta ) = \! \int _{0} ^{\infty} \!\! dk \frac{2Ze \epsilon _{1} \alpha _{-} J _{1} \left(k \rho \right) e ^{-k(z+2z _{0})}}{\epsilon _{1} \! \left( \alpha _{+} ^{2} + \alpha _{-} ^{2} \right) + \epsilon _{2} k ^{2} +  k  \alpha _{+} \! \left( \epsilon _{1} \! + \! \epsilon _{2} \right)} . \label{Eq:A}
\end{align}
This vector potential cannot be interpreted in terms of images, as in the case of a topological insulator, for which the magnetic field is due to an image magnetic monopole. As shown in Ref. \cite{PhysRevB.99.155142}, Eq.~(\ref{Eq:A}) can be interpreted in terms of an infinite number of sheets, one for each value of $z$ in the bulk, all supporting a surface Hall effect.

The natural geometry of the problem is provided by the prolate spheroidal coordinates $(\xi,\eta,\phi)$ which are related with the Cartesian coordinates as follows:
\begin{align}
    x &= z _{0} \sqrt{(\xi ^{2} - 1)(1-\eta ^{2})} \cos\phi , \notag \\ y &= z _{0} \sqrt{(\xi ^{2} - 1)(1-\eta ^{2})} \sin \phi , \notag \\ z &= z _{0} \eta \xi , 
\end{align}
with the range of the parameters given by
\begin{align}
    1 \leq  \xi \leq \infty ,\qquad -1 \leq \eta \leq 1 ,\qquad 0 \leq \phi \leq 2 \pi .
\end{align}

The usefulness of this coordinate system is apparent, since the plane $z = 0$, i.e. the interface between the WSM and the dielectric semiconductor or vacuum, is defined by the surface $\eta = 0$; hence, we restrict our calculations to the range $0 \leq \eta \leq 1$.

In the new coordinate system, the  potential energy (\ref{PotEnergy2}) takes the form:
\begin{align}
    V (\xi,\eta) &= -\frac{ e}{\epsilon _{2}  z _{0}} \left[ \frac{ Z }{ \xi - \eta }+\frac{ \epsilon _{2} - \epsilon _{1}}{\epsilon _{2} + \epsilon _{1} } \left(\frac{Z}{\left|\xi + \eta \right|} - \frac{1}{4 \eta \xi} \right)\right]\nn \\ & \phantom{=} + \frac{ e \epsilon _{1} }{\epsilon _{1} + \epsilon _{2}} \, \mathcal{I}(\xi,\eta) , \label{Eq:Phi_prolatas}
\end{align}
where
\begin{widetext}

\begin{align}
     \mathcal{I}(\xi,\eta) = \frac{1}{z _{0}} \int _{0} ^{\infty} dk \frac{ ( \gamma _{+} ^{2} + \gamma _{-} ^{2} - k ^{2} ) \left[ 2 Z J _{0} \left(k \sqrt{(\xi ^{2} - 1) ( 1 - \eta ^{2})} \right) e^{- k (\eta \xi + 1)} - e ^{-2 k \eta \xi } \right]}{\epsilon _{1} (\gamma _{+} ^{2} + \gamma _{-} ^{2}) + \epsilon _{2} k ^{2} +  k \gamma _{+} (\epsilon _{1} + \epsilon _{2} ) }, 
\end{align}
and the vector potential now becomes:

\begin{align}
    A _{Ze} (\xi,\eta) =  \int _{0} ^{\infty}  dk \frac{2 Z e \epsilon _ {1} \gamma _{-} J _{1} \left( k \sqrt{(\xi ^{2} - 1)(1 - \eta ^{2})} \right) e ^{- k (\eta \xi + 1)}}{\epsilon _{1} ( \gamma _{+} ^{2} + \gamma _{-} ^{2} ) + \epsilon _{2} k ^{2} + k \gamma _{+} (\epsilon _{1} + \epsilon _{2}) }. \label{Eq:A_prolatas}  
\end{align}

\end{widetext}
where now
\begin{align}
\gamma _{\pm} ( k ) = \sqrt{\frac{k}{2} \left( \sqrt{k ^{2} + \Lambda ^{2}} \pm k \right)} , \label{gamma}
\end{align}
being $\Lambda = z _{0} \Sigma$ a dimensionless parameter.

We observe that both the potential energy (\ref{Eq:Phi_prolatas}) and the vector potential (\ref{Eq:A_prolatas}) do not depend on the azimuthal angle. {One can notice from the potential (\ref{Eq:Phi_prolatas}) that the effective potential from the interface between both materials is attractive or repulsive depending on the sign of $(\epsilon_2-\epsilon_1)$ with effective charge $e/ z_0$.} In addition, given the Laplacian associated to the prolate coordinate system
\begin{align}
    \nabla ^{2} &= \frac{1}{z _{0} ^{2} \left(\xi ^{2} - \eta ^{2}\right)} \left[\frac{\partial}{\partial \xi} \left(\left(\xi ^{2} - 1 \right) \frac{\partial}{\partial \xi}\right)\right.\\
& \phantom{=} + \left.\frac{\partial}{\partial \eta}\left(\left(1-\eta^{2}\right) \frac{\partial}{\partial \eta}\right)+\frac{\xi^{2}-\eta^{2}}{\left(\xi^{2}-1\right)\left(1-\eta^{2}\right)} \frac{\partial^{2}}{\partial \phi^{2}}\right], \nn
\end{align}
it is clear that the full system (WSM+atom) has azimuthal invariance. Therefore,
the wave function can be separated as
\begin{align}
    \Psi _{nm} (\xi,\eta,\phi) = \psi _{nm} (\xi,\eta) \frac{e^{\mathrm{i}m\phi}}{\sqrt{2\pi}}, \label{Ansatz}
\end{align}
where the constant $m$ must be an integer for the wave function to be a single-valued and $n$ is additional quantum number to be determined. Applying the quantum Hamiltonian (\ref{Hamiltonian}) to the wave function (\ref{Ansatz}) we get the eigenvalue equation
\begin{align}
    \left[-\frac{\hbar ^{2}}{2 \mu}\nabla ^{2} + V _\text{eff} ^{m} (\xi,\eta) \right] \psi  _{nm} (\xi,\eta) = \mathcal{E} _{m} \psi_{nm} (\xi,\eta),
\end{align}
where $\mathcal{E}_m$ is the energy of the state with quantum number $m$ and $V_\text{eff}^m(\xi,\eta)$ is the effective potential given by:
\begin{align}
    V _\text{eff} ^{m} (\xi,\eta) &= V (\xi,\eta) + \frac{m ^{2} \hbar ^{2}}{2 \mu z _{0} ^{2}(\xi ^{2} - 1)(1 - \eta ^{2})}  \nn \\ & \phantom{=} + \frac{e ^{2}}{2 \mu c ^{2}} A _{Ze} ^{2} (\xi,\eta) + \frac{ me \hbar A _{Ze} (\xi,\eta)}{z _{0} \mu c \sqrt{(\xi ^{2} - 1)(1 - \eta ^{2})}} . \label{Eq:Veff_prolatas}
\end{align}
The first term corresponds to the electrostatic interactions (including the one coming from the anomalous Hall effect), the second term is the usual centrifugal potential, while the third and forth terms are the diamagnetic and paramagetic components of the Hamiltonian, respectively.

\section{Variational calculus}\label{sec:Variational_Functions}
In order to compute the energies of the lowest states, we used the standard variational method. We follow a recipe for choosing trial functions based on the product of $1s$ Slater orbitals in order to reproduce the correct behavior near the Coulombic singularities. This recipe to design compact wave functions has been widely applied for studying atoms, molecules, and quantum dots~\cite{Turbiner:1984,Turbiner2006309,artturbiner,NaderPhysRevA.100.012508,Nader2017}. Thus, the trial function is a product of three Slater orbitals corresponding to each Coulombic interaction:
electron-nucleus, electron-image of the nucleus, and electron-image of the electron, i.e.
\begin{eqnarray}
\psi_0 &=& \eta e^{-\alpha_1 r_1-\alpha_2 r_2 + \alpha_c r} 
\nonumber \\ &=&\eta e^{-\alpha_1(\xi+\eta)}e^{-\alpha_2(\xi-\eta)}e^{2\alpha_cz_0\xi\eta},
\label{psi0}
\end{eqnarray}
where $\alpha_1,\alpha_2$ and $\alpha_c$ are variational parameters and the factor $\eta$ is introduced in order to satisfy the boundary condition $\psi(z=0)=0$.
In general, for any state, we consider a wave function in the form of Eq.~(\ref{psi0}) (with its own set of variational parameters) multiplied by a convenient factor that guarantees orthogonality:
\bea
\label{wfes}
\psi_{nm}({\bf r})= \left[\left(\xi^{2}-1\right)\left(1-\eta^{2}\right)\right]^{|m|/2}\psi_0 f_{n}(\eta,\xi)\frac{e^{\mathrm{i}m\phi}}{\sqrt{2\pi}},
\label{anzats2}\nn\\
\eea
{ where $f_n(\eta,\xi)$ is a polynomial of degree $n$ which in turns, indicates the number of radial nodes of the wave function. Due the numeric nature of our solutions, notice that $n$ does not match the principal quantum number of the Hydrogen atom where the radial nodes of the wave function are given by $n-l-1$. Here, $n$ serves as a label for the radial trial function. Therefore, states with angular momentum $m>0$ represent angular excitations, while $n>0$ stands for radial excitations.}
At large distances $z_0$, the wave function of Eq.~(\ref{wfes}) goes to the $1s$ and $2p$ orbitals of the Hydrogen atom for ($n=0,m=0$) and ($n=0,m=1$), respectively. 
In particular, for the excited state $n=1,m=0$ we use
\bea
\label{factor}
f_1(\eta,\xi)=(\alpha-r_2) , 
\eea
where $r_2$ is the distance from the electron to the nucleus and $\alpha$ is a constant.
With this form of the polynomial (\ref{factor}), the wave function (\ref{wfes}) reproduce the $2s$ orbital of the hydrogen atom at large distances $z_0$.
The parameter $\alpha$ is chosen such that it guarantees orthogonality between the excited state $n=1,m=0$ and the ground state $n=0, m=0$ 
\bea
\alpha=\frac{\int d^3\mathbf{r}~r_2 \psi_0(\mathbf{r})\psi_{10}(\mathbf{r})}{\int d^3\mathbf{r}~ \psi_0(\mathbf{r})\psi_{10}(\mathbf{r}) }.
\eea

It is worthy to mention that the wave function of Eq.~(\ref{psi0})
is square normalizable as long as the argument in the exponential remains positive. Since the effective potential of Eq.~(\ref{Eq:Phi_prolatas}) can be repulsive, if the ratio $e/z_0$ is large enough the system can be ionized and therefore the wave function losses square normalizability i.e. a critical effective charge $e/z_0$ appears. The estimate of critical charges is a very active area in atomic systems~\cite{critical2,critical3,critical4,critical5,critical6}, and in this work such a feature will impact the allowed transitions of the system, as is discussed in Secs.~\ref{sec:The ground state} and~\ref{sec:The excited states}.

We evaluated the two involved integrals (numerator and denominator) numerically in cylindrical coordinates, 2-dimensional and 3-dimensional, including topological effects, employing the adaptive multidimensional routine \cite{GENZ1980295}. For each integral, the integration space was subdivided into 6 (12 including topological term), in which the integration was done separately.
The partitioning is adjusted and controlled depending on the variational parameters and the value of $z_0$.
 We performed the minimization in the parameter space by using the subroutine MINUIT of CERN-LIB~\cite{JAMES1975343}. The code used in this work is an adaptation of the FORTRAN code designed by A. V. Turbiner and J. C. Lopez Vieyra.
 
for the variational calculus of atoms and molecules in strong magnetic fields \cite{Turbiner2006309}. .

\section{Results and Discussion}\label{sec:Results_and_Discussion}

In order to elucidate the impact of the anomalous Hall effect, a hallmark of the Weyl semimetallic phase, upon a hydrogen-like atom near of its surface, we study three experimentally-accessible configurations: a single hydrogen atom in vacuum, and hydrogenic GaAs {and GaP} impurities. The first one can be achieved by placing a non-interacting hydrogen gas close to the surface of the WSM. The second is accomplished with semiconductor growing techniques, which are implemented in thin films. For an hydrogen in vacuum we take $\epsilon _{2}=\epsilon_0=1$ and the electron's mass $m_{0}$. For the GaAs impurity, we use $\epsilon_2=13.18$ for the relative permittivity and $\mu=0.067m_0$ for the effective mass, { while for the GaP impurity we use $\epsilon_2=9.1$ for the relative permittivity and $\mu=0.35 m_0$ for the effective mass~\cite{HandbookSC}.}

\begin{figure}[h!]
    \centering
    \includegraphics[scale=0.5]{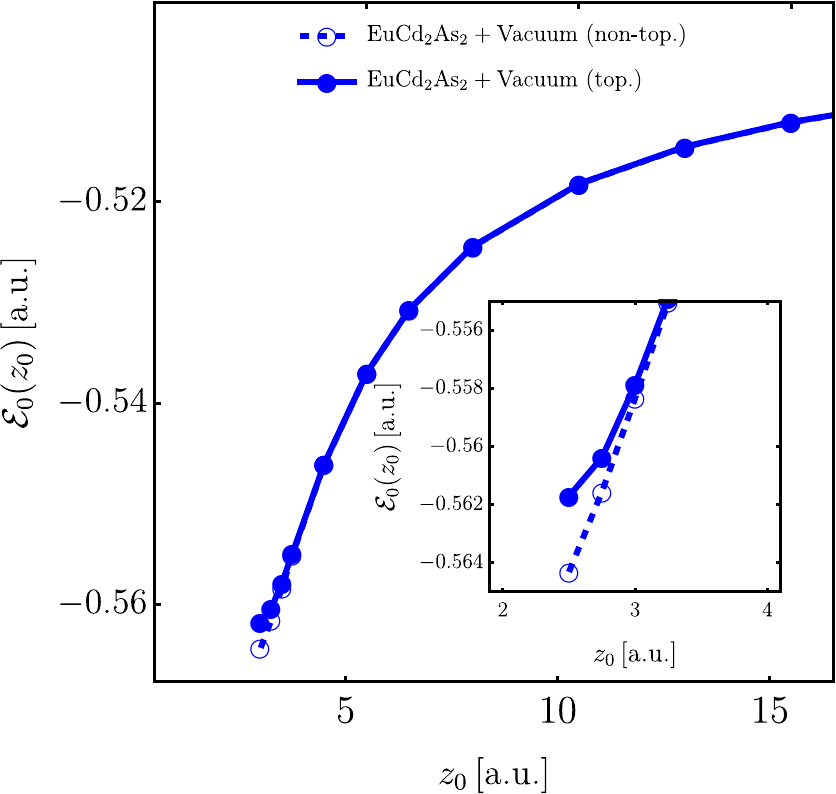}
    \caption{Ground state energy $\mathcal{E}_0$ for an hydrogen atom in vacuum ($\epsilon_2=1$) close to the WSM EuCd$_2$As$_2$ ($\epsilon_1=6.2$) as a function of the distance $z _{0}$ (in atomic units a.u.). The dashed line-open symbols is the calculation performed without topological terms, whereas the continuous line-filled symbols is the result with topology included.}
    \label{Fig:E0vacuum}
\end{figure}


{
For the Weyl semimetal sample we use EuCd$_2$As$_2$. This material hosts a single pair of Weyl nodes located at ${\boldsymbol{k}}=(0,0,\pm 0.03) 2 \pi /c $ at the Fermi level when the Eu spins are fully aligned along the $c$ axis (here $c = 0.729$nm, such that $b \sim 5.1 \times 10 ^{8}$m${}^{-1}$) \cite{PhysRevB.100.201102}. In this case the anomalous Hall effect is fully described by the theory introduced in Sec. \ref{sec:Electromagnetic_response_of_WSM} and hence the analysis of the atom-WSM interaction of Sec. \ref{sec:Hydrogenlike_ion_near_the_WSM_surface} is applicable, since the surface that does not support Fermi-arc electronic states is properly identified (in this case is the $xy$-plane since ${\bf{b}} = b \hat{{\bf{e}}}_{z}$). {The data, including energies and optimal variational parameters obtained in this work, can be  found in the repository linked in Ref.~\cite{repositorio}.}


}

\begin{figure}
    \centering
        \includegraphics[scale=0.5]{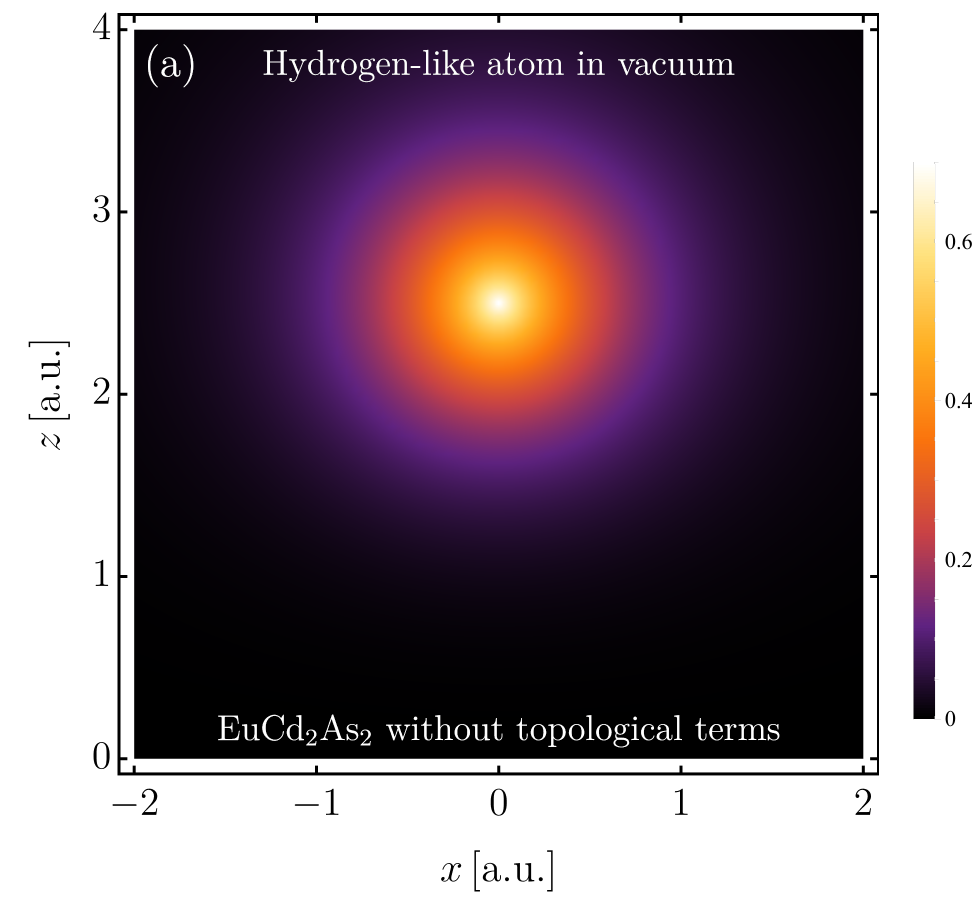}\\
        \vspace{0.3cm}
    \includegraphics[scale=0.5]{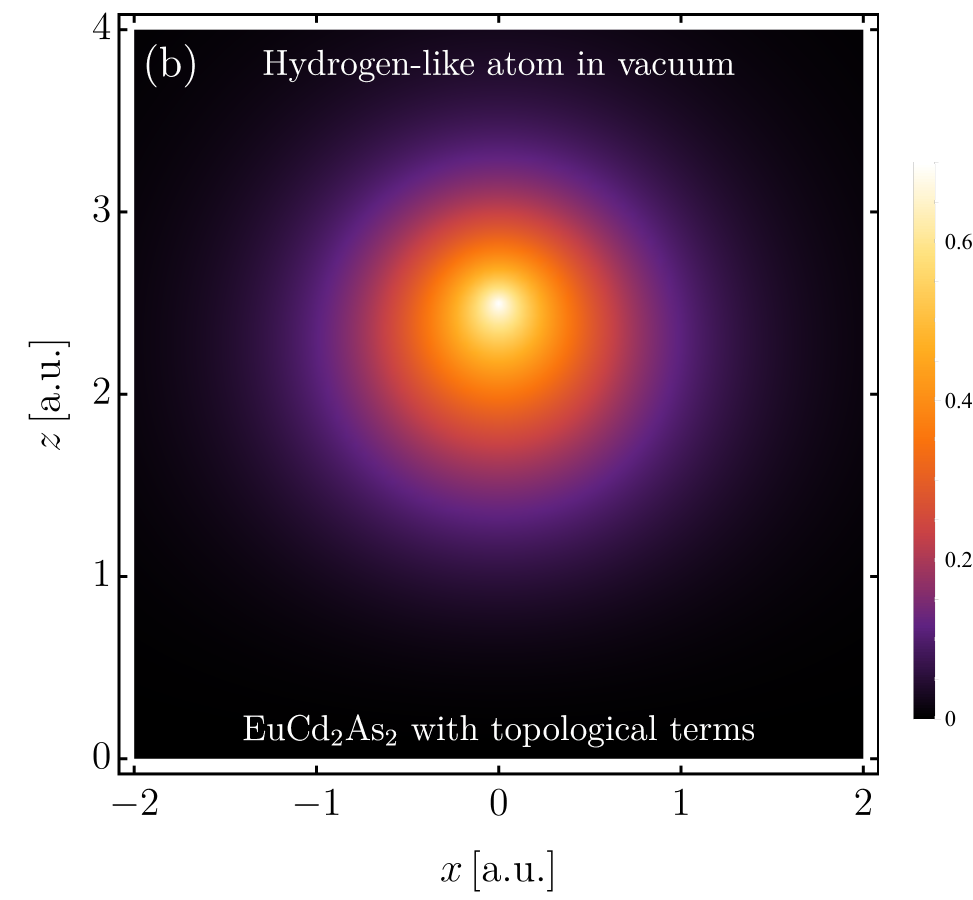}
    \caption{Probability density function for the ground state of the hydrogen atom in vacuum located at $z_0=2.5$ a.u. from EuCd$_2$As$_2$ surface: (a) without topological effects and (b) with topological terms.}
    \label{fig:densityTaAsVacuum}
\end{figure}

\subsection{The ground state}\label{sec:The ground state}

Figure~\ref{Fig:E0vacuum} shows the ground state energy $\mathcal{E}_0$ as a function of the distance $z_0$ for an hydrogen atom in vacuum located near to the WSM {EuCd$_2$As$_2$}, for which $\epsilon_1=6.2$ \cite{chang2012topological}. As we can see, the ground state has a shift in energy as compared with the nontopological case when the atom is close to the surface. As expected, at large distances, the interaction with the surface is negligible, and hence the results converge to the standard description of the hydrogen atom in the ground state. { In this case, for the sake of clarity, we plot the energy including distances of a few atomic radii, where we know the electromagnetic response theory of Sec. \ref{ModelHamiltonian} ceases to be valid. In a similar fashion, in}  Fig.~\ref{fig:densityTaAsVacuum} we plot the probability density for the system. We observe that in the absence of the topological term (i.e. for $\sigma _{xy}=0$), the electronic cloud is repelled by the material, and as we can directly read from Eq. (\ref{PotEnergy2}), this is due to the repulsive Coulomb-like interaction between the orbiting electron in vacuum and its image charge within the sample given that $\epsilon_{2}<\epsilon_{1}$ for a vacuum-{EuCd$_2$As$_2$} junction. However, as Fig.~\ref{fig:densityTaAsVacuum}-(b) indicates, the effect of the WSM's nontrivial topology is to compensate such repulsion so that the electronic cloud is attracted to the wall. Physically, being positive the image charge, the atomic electron is always followed by its image charge, and these atomic currents attracts each other according to the Ampere's force law.

\begin{figure}
    \centering
    \includegraphics[scale=0.5]{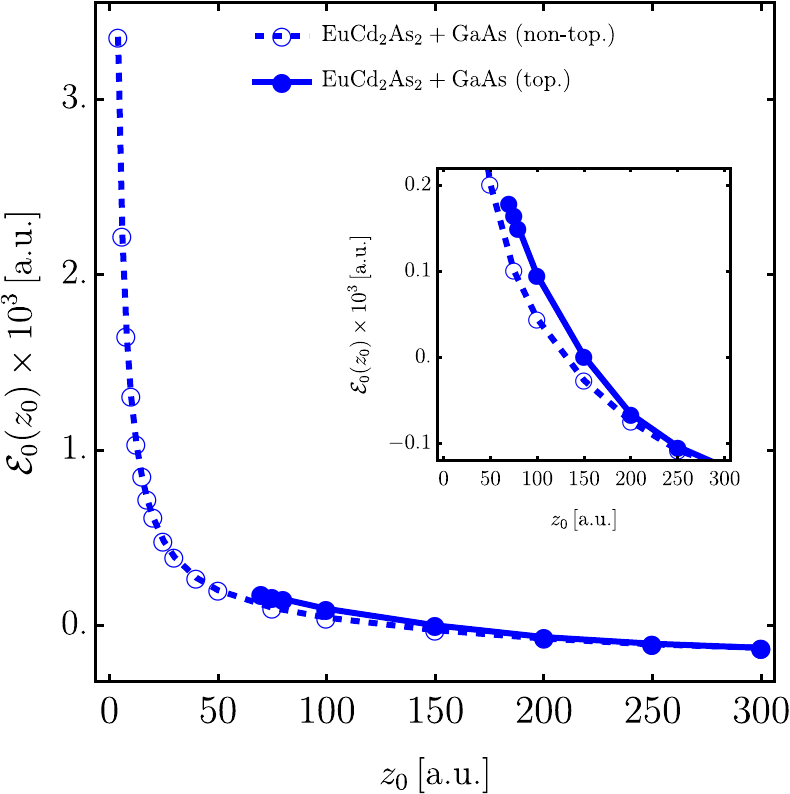}
    \caption{Ground state energy $\mathcal{E}_0$ (scaled by a factor of $10^3$) for a hydrogenic impurity in GaAs ($\epsilon_2=13.18$) close to the WSMs EuCd$_2$As$_2$ ( $\epsilon_1=6.2$)  as a function of the distance $z _{0}$ (in atomic units a.u.). The dashed line-open symbols line is the calculation performed without topological terms, whereas the continuous line-filled symbols is the result with topology included.}
    \label{Fig:E0withMaterials}
\end{figure}

\begin{figure}
    \centering
    \includegraphics[scale=0.5]{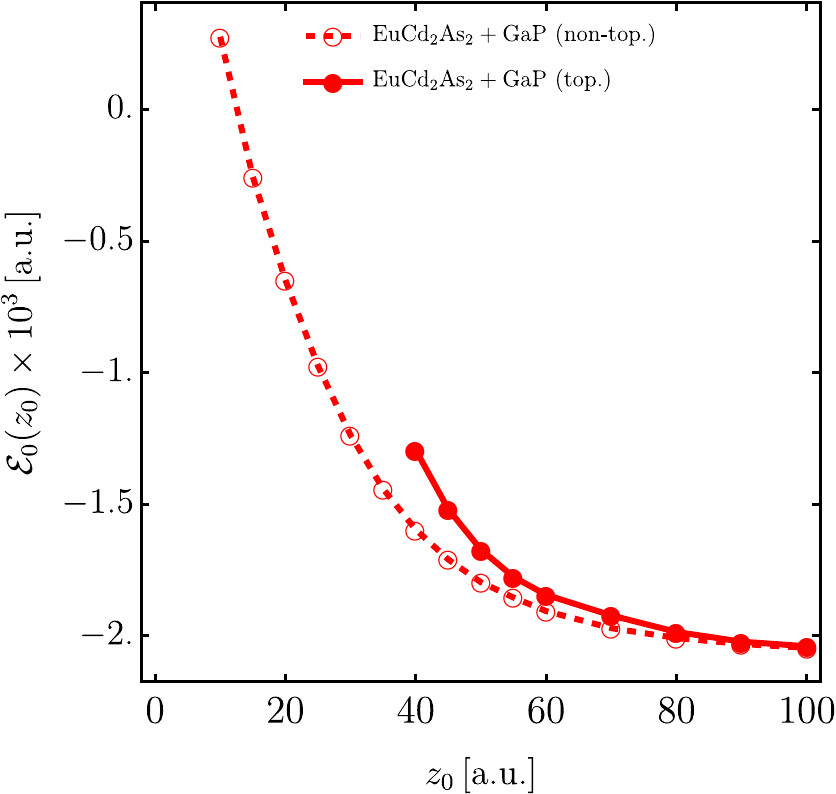}
    \caption{Ground state energy $\mathcal{E}_0$ for a hydrogenic impurity in GaP ($\epsilon_2=9.1$) close to the WSMs EuCd$_2$As$_2$ ( $\epsilon_1=6.2$)  as a function of the distance $z _{0}$ (in atomic units a.u.). The dashed line-open symbols line is the calculation performed without topological terms, whereas the continuous line-filled symbols is the result with topology included.}
    \label{Fig:E0withMaterials2}
\end{figure}

\begin{figure}
    \centering
        \includegraphics[scale=0.5]{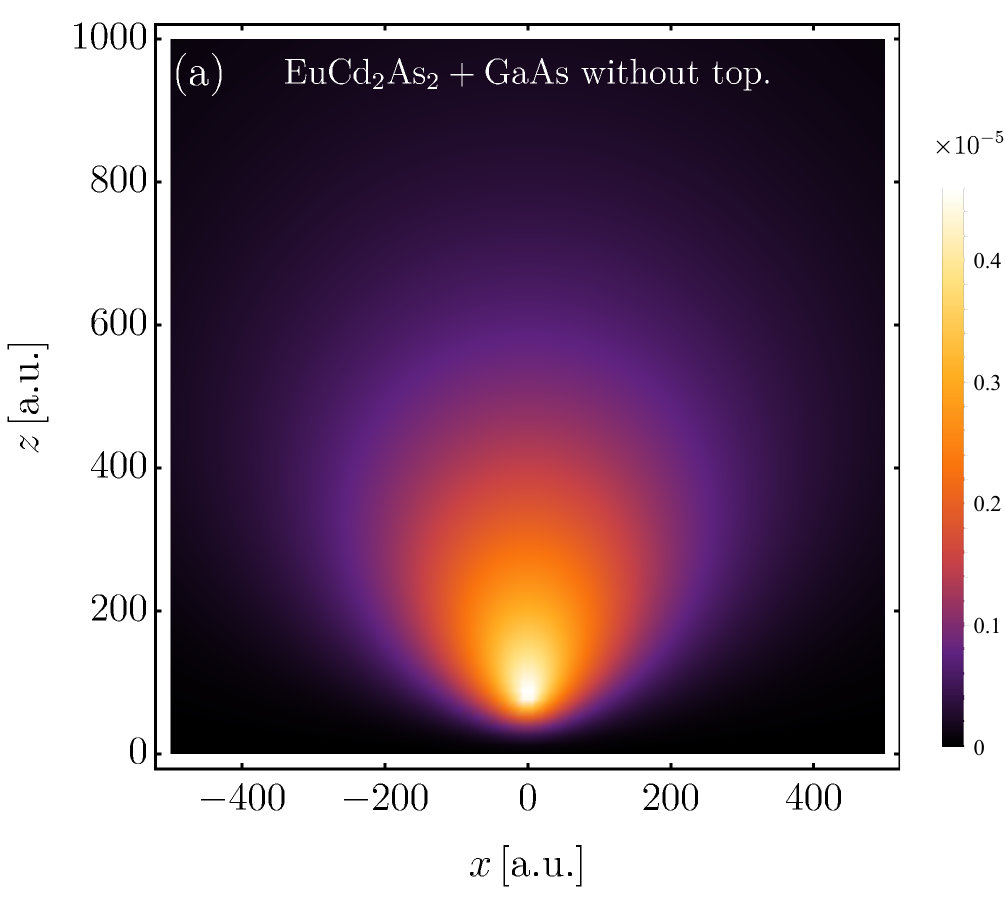}\\
        \vspace{0.3cm}
        \includegraphics[scale=0.5]{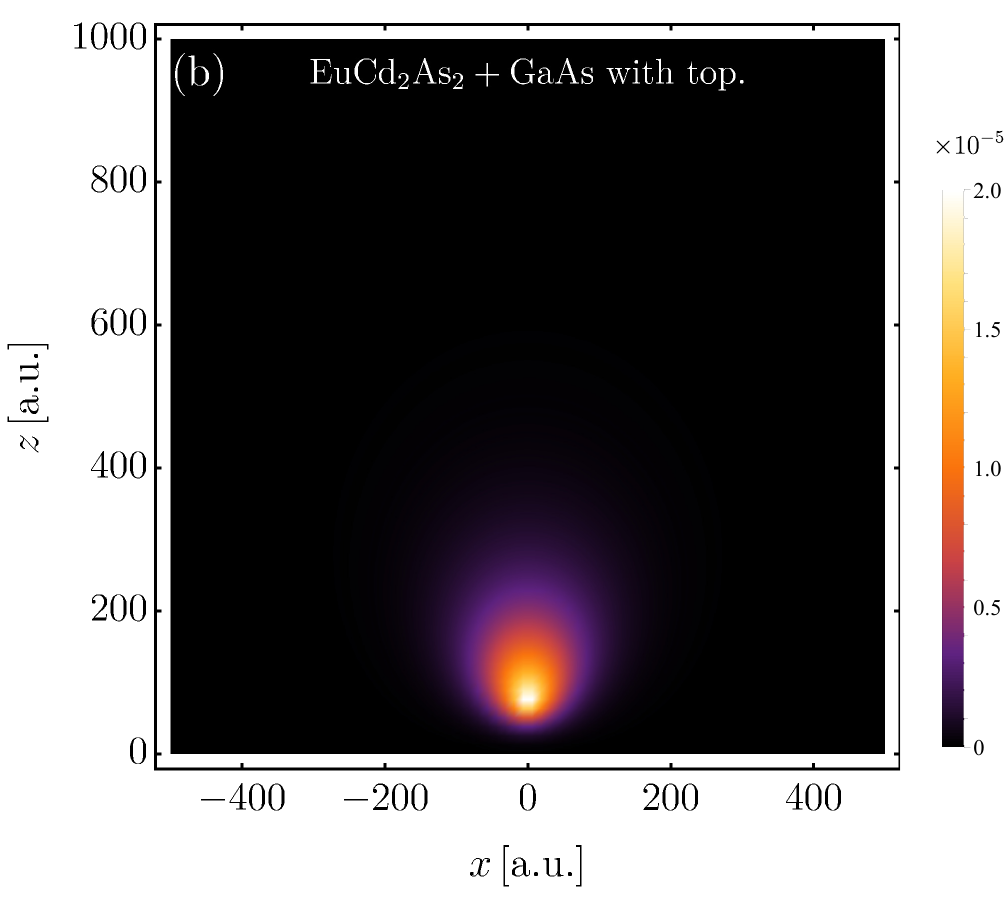}
    \caption{Probability density function for the ground state for the junction EuCd$_2$As$_2$ + GaAs (a) without topological terms, and (b) with topological terms both for $z_0=75$ a.u.}
    \label{fig:densityTaAsGaAsTop}
\end{figure}

\begin{figure}[h!]
    \centering
       \includegraphics[scale=0.5]{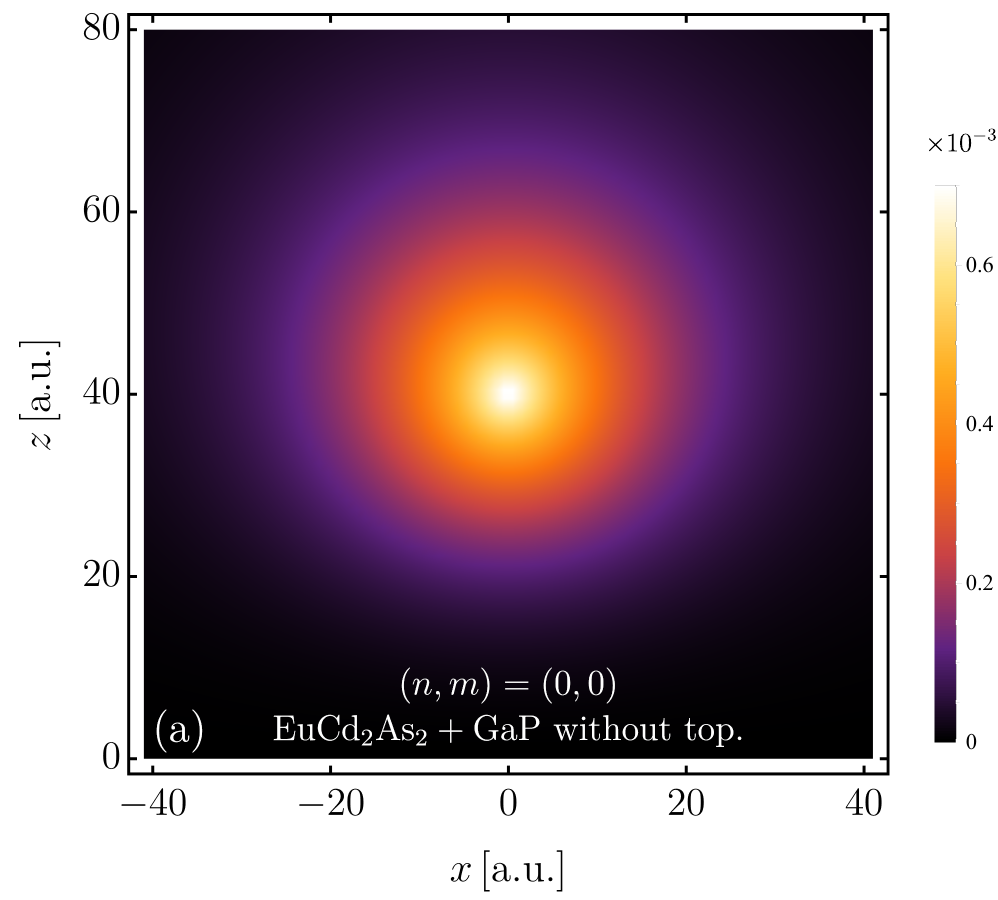}\\
       \vspace{0.3cm}
    \includegraphics[scale=0.5]{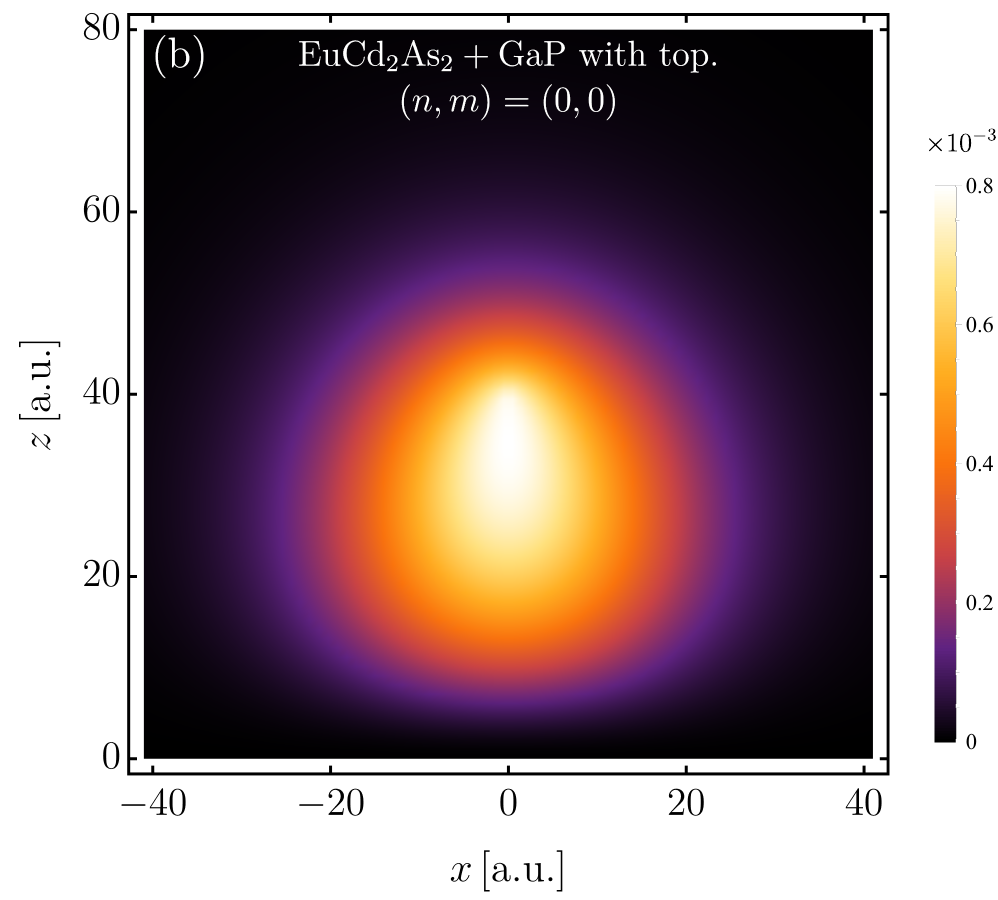}
    \caption{Probability density function for the ground state with $z_0=40$ a.u}
    \label{fig:densityTaAsGaPground}
\end{figure}

\begin{figure*}
    \centering
    \vspace{0.3cm}
    \includegraphics[scale=0.5]{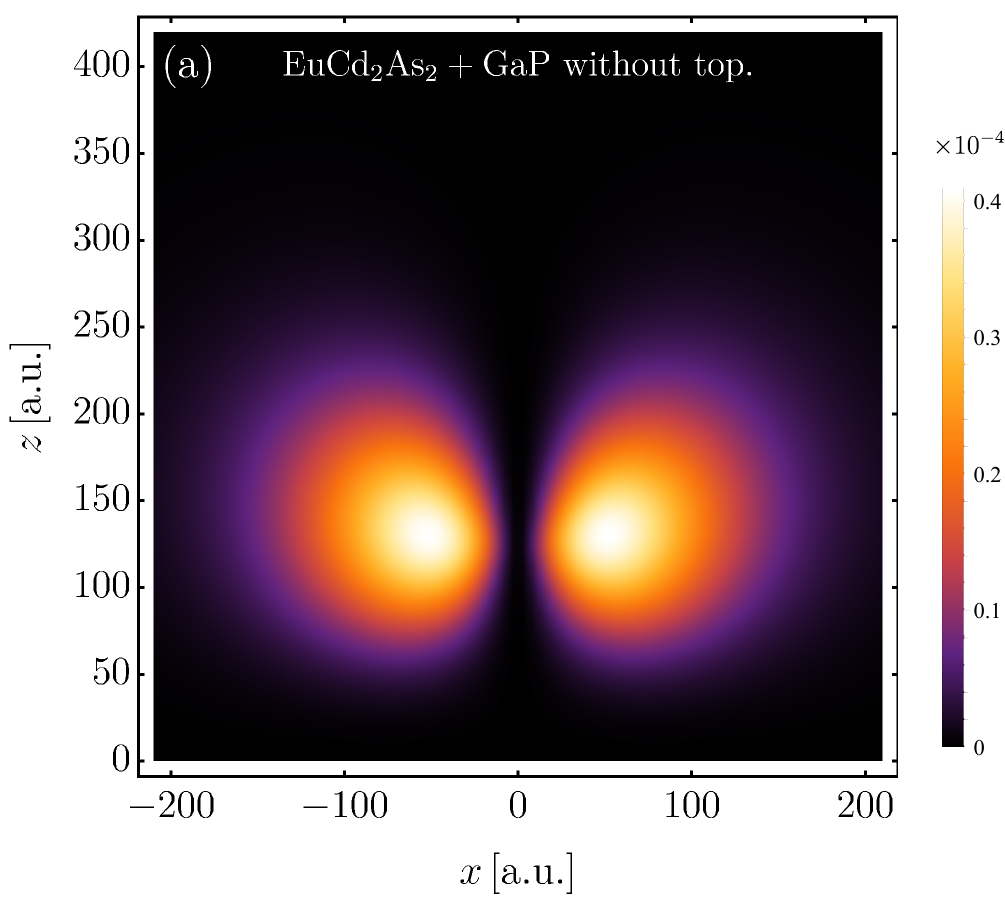}\hspace{0.3cm}
    \includegraphics[scale=0.5]{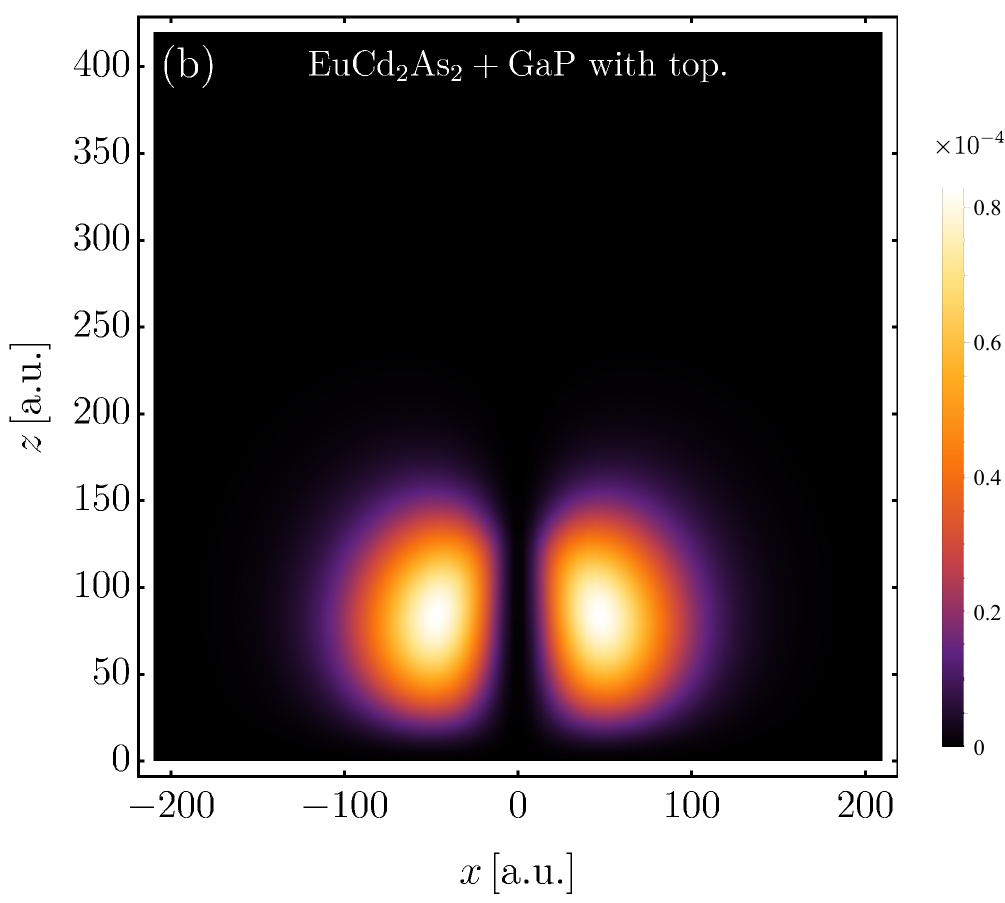}\\
    \vspace{0.3cm}
     \includegraphics[scale=0.5]{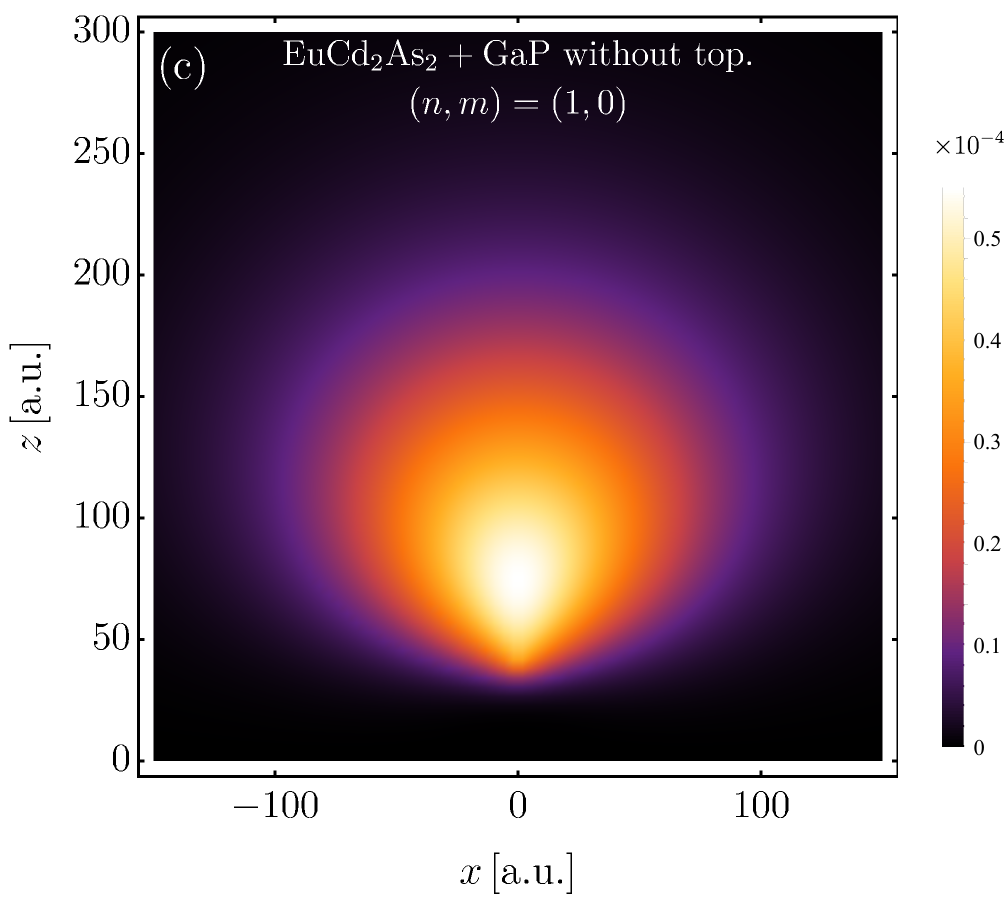}\hspace{0.3cm}
    \includegraphics[scale=0.5]{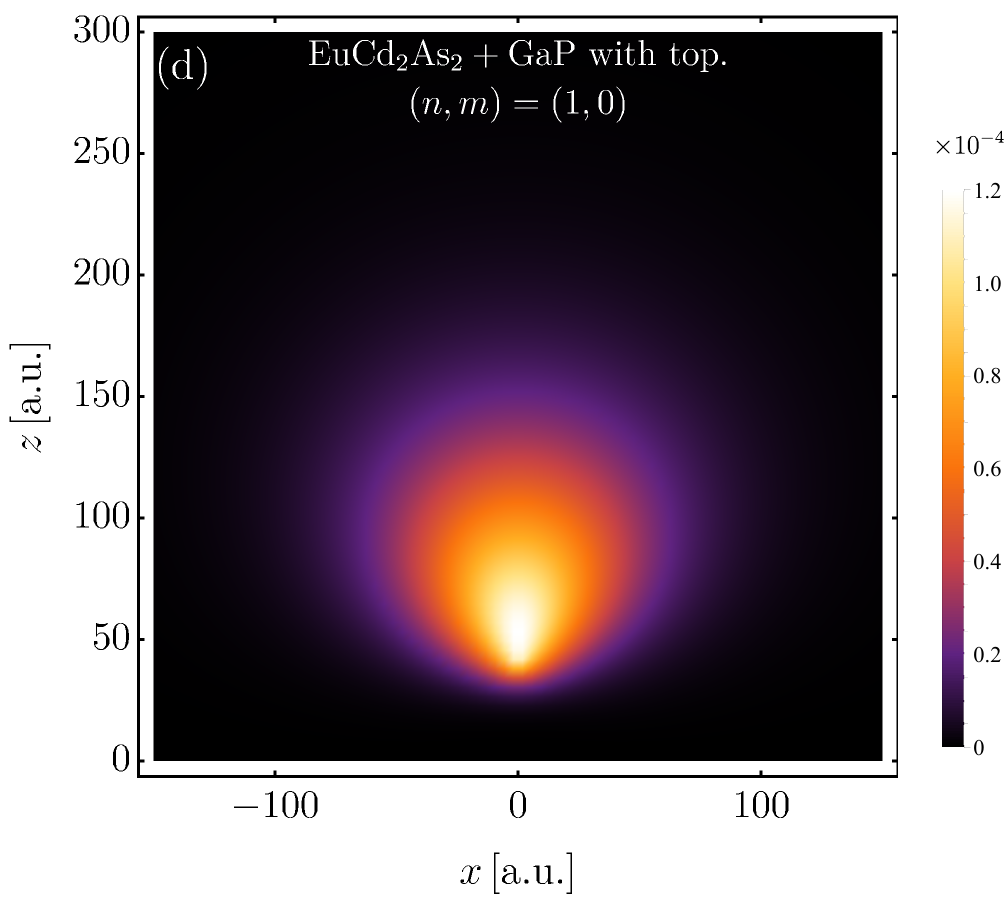}
    \caption{Probability density function for the excited states $(n,m)= (0,1)$, and $(1,0)$, for the junction EuCd$_2$As$_2$ + GaP. In each panel, the hydrongen-like ion is located at: (a)-(b) $z_0=125$ a.u, and (c)-(d) $z_0=40$ a.u.}
    \label{fig:densityTaAsGaPexcited}
\end{figure*}

The second possibility we shall consider is an hydrogen-like impurity embedded in GaAs close to a Weyl semimetal. As we can see from Eq. (\ref{Eq:Phi_prolatas}), the sign of the interaction potential can be tuned by means of the permittivities. For an hydrogen atom in vacuum it is clear that {$ \delta \epsilon \equiv \epsilon _{2} - \epsilon _{1} < 0$}; however, a GaAs has a larger permittivity such that $\delta \epsilon$ flips its sign. Figure~\ref{Fig:E0withMaterials} shows the ground state of a hydrogen-like impurity embedded in GaAs and close to the WSM EuCd$_2$As$_2$. When the topological contribution is considered, the last term in the interaction potential of Eq.~(\ref{Eq:Phi_prolatas}), which is positive definite, overwhelm the second term and hence the resulting ground state energy becomes positive for small $z_0$ and tends to zero as the atom moves far away from the surface. In order to understand this effect, in Fig.~\ref{fig:densityTaAsGaAsTop} the probability density function for a GaAs impurity is plotted. This probability distribution shows that both cases (with and without topological terms) support bound states. However, the nontopological calculation describes an extended atomic cloud (due to the small effective mass and the large effective dielectric constant~\cite{kohn1957shallow}), which is strongly confined when the topological term is considered. Then, the presence of the WSM localizes the charge carriers which may affect the enhancement of electric conductivity in doped semiconductors.

{ A similar situation is observed in Fig.~\ref{fig:densityTaAsGaPground}, where we changed the semiconductor sample by GaP: by ignoring the topological terms the electronic cloud is slightly repelled by the materials interface, but when the topology is turned-on there is a strongly confinement and deformation of the spatial probability distribution. The latter is not only correlated with the value of the semiconductor's relative permittivity, but also with electron's effective mass ($\mu_\text{GaP}\gg \mu_\text{GaAs}$), so that if $\mu$ enhances then the topological effects are appreciable given that the kinetic term of Eq.~\eqref{Hamiltonian} is not dominant in the Hamiltonian.

It is worth to mention that, for the electromagnetic field theory we have considered to be valid, it is required that the atom-WSM distance be large as compared with the atomic radius. However, as one can see in Fig. \ref{fig:densityTaAsVacuum} for an hydrogen in vacuum near the WSM sample, the effects of the topological nontriviality are appreciable only when the atom is close enough to the surface, at a few atomic radii. Therefore, although interesting, this case is not realistic. For the topological effects to be observed in a region where our model applies we turned to use semiconductor samples, since the effective electron mass is smaller (than the electron mass) and hence the unperturbed atomic cloud is less confined. This situation opens the possibility to be furthest from the surface, where our model works, and observe anomalous Hall effect signals upon the atomic could and energy levels. This is confirmed by our plots of the ground state energy: while significant effects for a genuine hydrogen take place at the length scale of few atomic units (see Fig. \ref{Fig:E0vacuum}), they appear in a farther region for semiconducting samples.
}


We have to point out that, to construct Fig.~\ref{fig:densityTaAsGaAsTop}, the value of $z_{0}$ is not arbitrary at all. Indeed, for enough small values of $z_{0}$, our variational method to solve the Schr\"{o}dinger's equation produces a not normalizable wave function. To better understand this fact we have to study the shape of the effective potential of Eq.~(\ref{Eq:Phi_prolatas}). On the one hand, we observe that due to the Coulombic term, the effective electrostatic potential exhibits an infinite potential well at the origin, thus supporting bound state solutions. On the other hand, it is clear from Eq. (\ref{Eq:A}) that the vector potential vanishes for $\rho = 0$, i.e. along the line perpendicular to the interface.  Also, it is clear that the potential diverges at the surface, and this is due to the divergent electron-image electron interaction there. Therefore, a critical point $z _{c}$ must exist between the surface and the position of the nucleus, i.e. $-1 < z _{c} / z _{0} < 0$. For $z _{0} > z _{c}$, bound state solutions are allowed, and hence the corresponding wave functions are normalizable, as they should be. However, when the nucleus-surface distance is below the critical point, $z _{0} < z_{c}$, there are no bound states solutions since the atom becomes ionized, and consequently the wave function stop to be normalizable, as evinced by our numerical solutions.
For the permittivity configuration of EuCd$_2$As$_2$+vacuum Fig.~\ref{Fig:E0vacuum} the critical value is $z_c\sim2.5$ a.u.. {A detailed discussion about the existence of $z_c$ is given in Appendix~\ref{Ap:zc}.}

\begin{figure}[h!]
    \centering
    \includegraphics[scale=0.5]{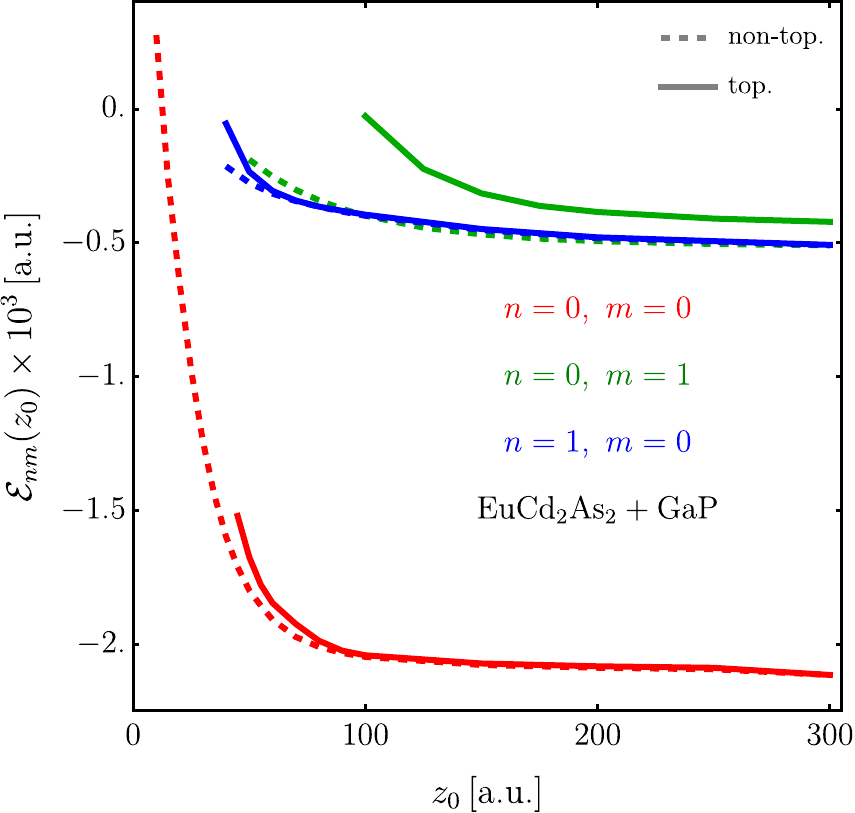}
    \caption{ Ground and excited states for an hydrogen-like atom in GaP ($\epsilon_r=9.1$) for the cases with and without topological terms in the WSM EuCd$_2$As$_2$ as a function of the distance $z _{0}$ (in atomic units a.u.). Note the shift of $z_0^\text{critical}$ for $m=1$.}
    \label{Fig:ExcitedStates}
\end{figure}


\subsection{The excited states}\label{sec:The excited states}

{Figure~\ref{fig:densityTaAsGaPexcited} shows the probability density function of the excited states $n=1,m=0$ and $n=0,m=1$ for the juncture EuCd$_2$As$_2$+GaP. Moreover, Fig.~\ref{Fig:ExcitedStates} depicts the energy for those states as a function of the distance $z_0$ for an Hydrogen-like impurity in GaP near the WSM. As can be noticed, the attractive effect of the WSM is present even in the excited states, which implies a considerable deformation of the electron's probability cloud. In the case of Fig.~\ref{Fig:ExcitedStates}, each plot suddenly cutoff at a different critical distance. Indeed, the ground's state critical distance is smaller than the corresponding to excited states, as evinced in the figure. 
{ This means that there exist a certain range of distances $z_0$ where the ground state is bound but the excited states are not.}
At large distances each state goes asymptotically to the energy of the Hydrogen atom in GaP, namely
\bea
\lim_{z_0\to+\infty}\mathcal{E}_n=-\frac{\mu}{2n^2\epsilon_2^2},
\eea
with permittivity $\epsilon_2=9.1$ and effective mass $\mu=0.35$.

\subsection{Transition Amplitude}
In order to know the allowed transitions between the computed states, let us define the following transition amplitude in the dipole approximation:
\bea
\mathcal{M}_{(n_0,m_0)}^{(n,m)}\left(\hat{\mathbf{e}}_j\right)&\equiv&\bra{n,m;z_0}\hat{\mathbf{e}}_j\cdot\left(\mathbf{r}-\mathbf{r}_0\right)\ket{n_0,m_0;z_0},\nn\\
\label{selectionrules}
\eea
where $\hat{\mathbf{e}}_j$ is a unit vector that depends on the polarization of an external monochromatic light source. The defined quantity is related to the electron's probability to perform a transition which, by construction, depends on the ion location $\mathbf{r}_0=z_0\hat{\mathbf{e}}_z$.

\begin{figure}
    \centering
    \includegraphics[scale=0.5]{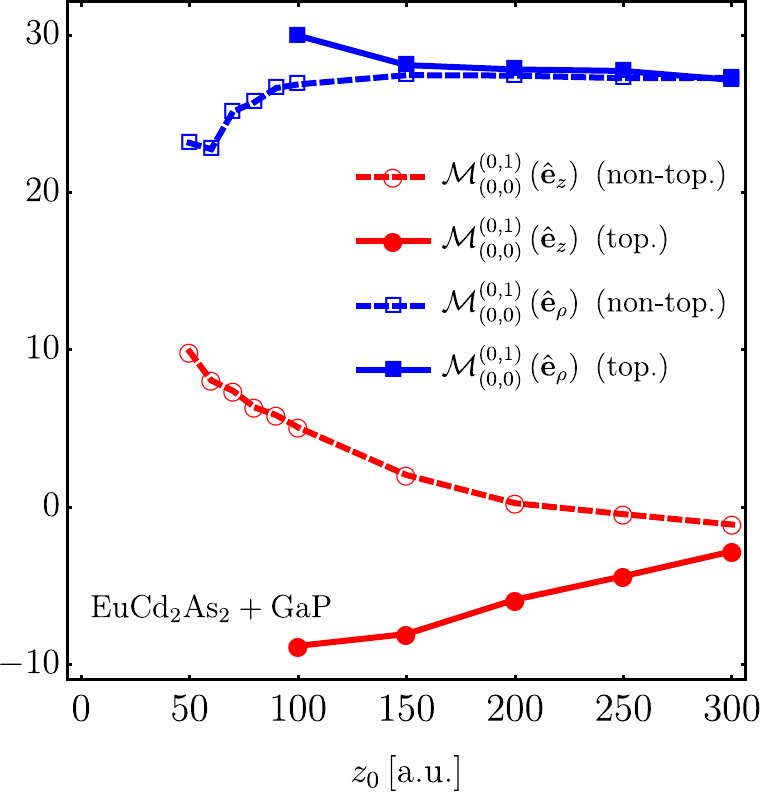}
    \caption{Transition amplitudes of Eq.~\eqref{selectionrules} for the transition $\ket{0,0}\to\ket{0,1}$.}
    \label{fig:SelectionRule_n0m0_n0m1}
\end{figure}

\begin{figure}
    \centering
    \includegraphics[scale=0.5]{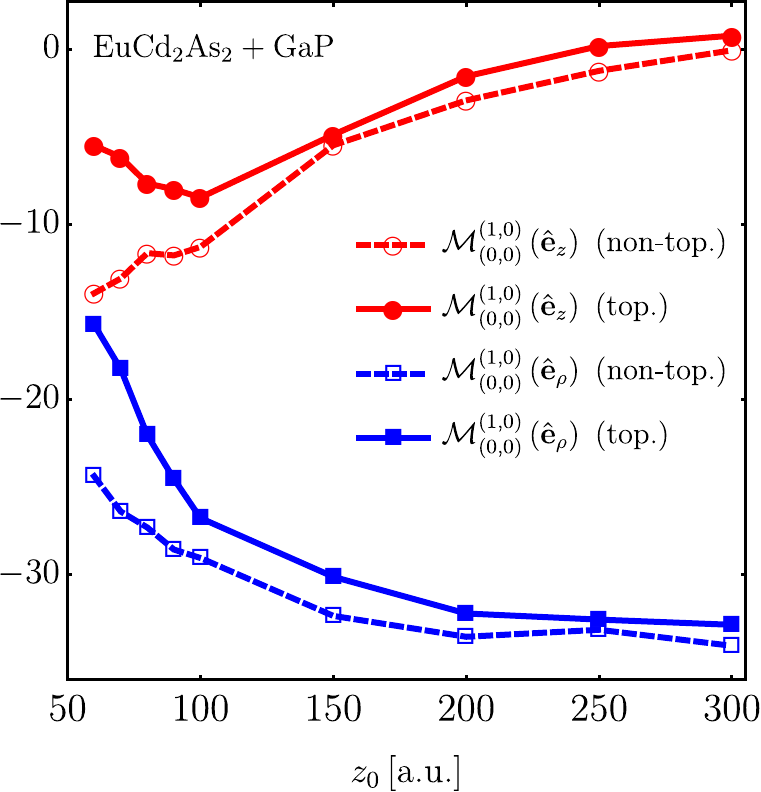}
    \caption{Transition amplitudes of Eq.~\eqref{selectionrules} for the transition $\ket{0,0}\to\ket{1,0}$.}
    \label{fig:SelectionRule_n0m0_n1m0}
\end{figure}

In the former discussion about the critical distance $z_0$ in which the wave function is still normalizable, one may wonder if the transitions from the ground to the excited states are available. For that sake, in Figs.~\ref{fig:SelectionRule_n0m0_n0m1} and~\ref{fig:SelectionRule_n0m0_n1m0} we present the transition amplitudes of Eq.~\eqref{selectionrules} for the transitions $\ket{0,0}\to\ket{0,1}$, and $\ket{0,0}\to\ket{1,0}$, respectively. The plots indicate that such transitions are allowed in the dipolar approximation, hinting that the impurity can perform the transition by changing its quantum numbers in distances up to the critical one, where we interpret that the ionization occurs. Also, the transition probability between states is appreciably modified when the topological effects are included, which can be used as a probe for testing the topological nature of the WSM. 
}


\subsection{The atomic polarization}

\begin{figure}
    \centering
    \includegraphics[scale=0.45]{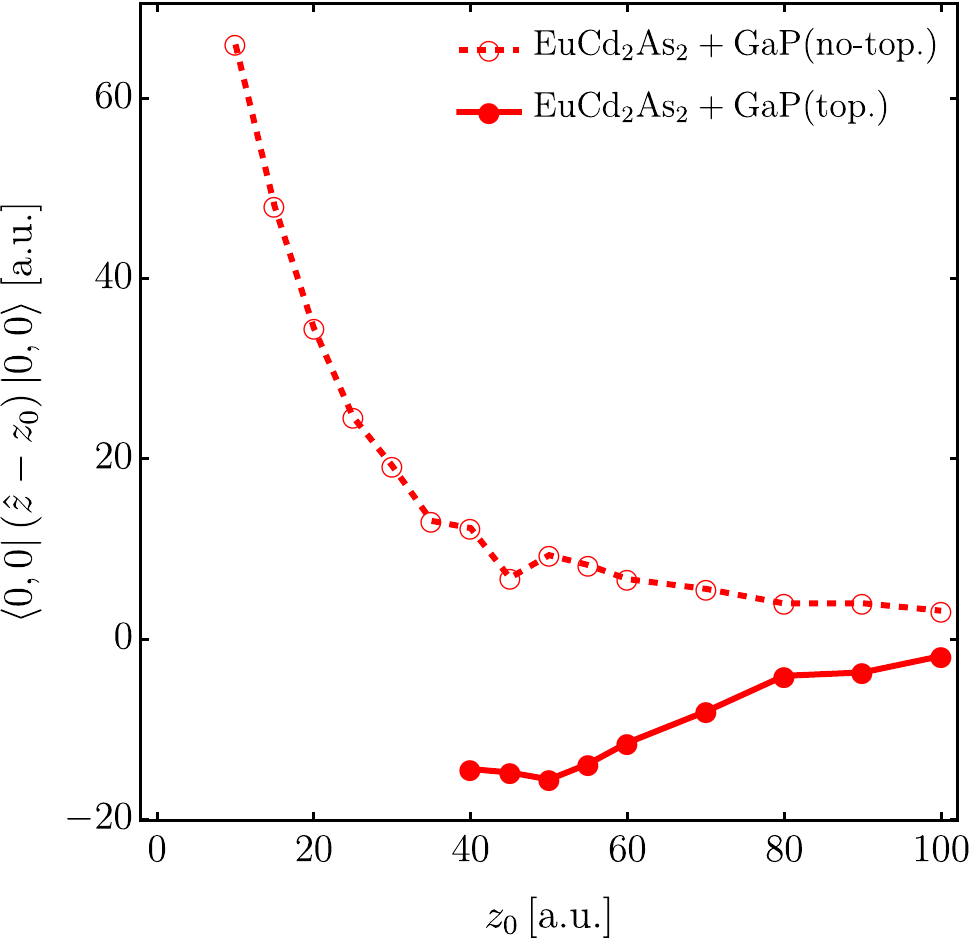}
    \caption{Dipolar matrix element $\langle 0,0|\left(\hat{z}-z_0\right)
    |0,0\rangle$ as a function of $z_0$ for a hydrogen atom in vacuum ($\epsilon_2=1$) close to the WSMs EuCd$_2$As$_2$. The dashed line-open symbols lines is the calculation performed without topological terms, whereas the continuous line-filled symbols is the result with topology included.}
    \label{figure:selectionruleforz}
\end{figure}
Finally, we study the effects of the WSM upon the atomic polarization. To this end, we use the standard definition of this quantity: $\langle 0,0|\left(\hat{z}-z_0\right)|0,0\rangle$. In Fig.~\ref{figure:selectionruleforz} we show the polarization for a hydrogenic GaP impurity close to the WSM EuCd$_2$As$_2$. The variational calculations show that the polarization points away from the surface, which we understand from the repulsion between the atomic cloud and its image. However, due the attractive nature of the topological sector in the effective potential, the polarization flips orientation for $\sigma _{xy} \neq 0$ and then points towards the surface. Also, as we can see in Fig.~\ref{figure:selectionruleforz}, the polarization decreases as increasing $z_0$, as expected, since the atom get back to be spherically symmetric and hence by parity considerations the polarization vanishes. On the other hand, for nucleus-surface distances $z_0$ below the critical distance $z_0$ the polarization is ill-defined due to the not normalizability of the wave function. The results have the same qualitative information if a hydrogen-like impurity in GaAs is considered but with an enhancement of $\langle 0,0|\left(\hat{z}-z_0\right)|0,0\rangle$, given the associated material parameters.


\section{summary and Conclusions} \label{sec:summary_and_Conclusions}

The interaction between atoms and surfaces has proven to be of fundamental importance in many branches of science like field theory, cosmology, molecular physics, colloid science, biology, astrophysics, micro- and nanotechnology, for example. The measurements of atom-surface interactions range from experiments based upon classical and quantum scattering, up to high precision spectroscopic measurements. Within the realm of atomic spectroscopy, hydrogenlike atoms provide an attractive test bed for studying new atom-surface interactions. It has been used, for example, to predict the effects of induced magnetic monopole fields in topological insulators upon the hyperfine structure of an hydrogenlike atom nearby \cite{PhysRevA.97.022502}. Following this idea, in this paper we consider the effects of a topological Weyl semimetal upon an hydrogenlike atom close to its surface. Importantly, unlike the TI case, the interaction between an atom and a metallic phase is rather stronger, and hence a perturbative analysis is not appropriate for small atom-surface distances.

In this paper we have analyzed the effects of a topological Weyl semimetal upon a hydrogen-like atom which is located in front of the face without surface states. Here we work in the nonretarded approximation, which is valid for atom-surface distances sufficiently large as compared with both the atomic radius and the distance between the atomic constituents of the sample. In this regime, the model Hamiltonian is based on the electromagnetic interaction between the atomic charges (atomic electron + nucleus) and the WSM. In the case of a dielectric, the interaction is computed as the Coulomb interaction between the atomic charges and the image electric charges. In a topological insulator, besides image electric charges, image magnetic charges appear as well, whose magnetic fields will in turn interact with the atomic electron (through the minimal coupling prescription). In the problem at hand, the electromagnetic interaction cannot be interpreted in terms of charge-image charge Coulomb interaction, since Maxwell equations are modified in the bulk of the WSM. However, assuming a WSM in the equilibrium state and at the neutrality point, the electromagnetic interaction can be modeled in an analytical fashion.

Using variational methods, we solve the corresponding Schr\"{o}dinger equation and determine the energy and wave function for the ground state of the system. This theory can be applied to two different configurations with possible experimental opportunities. On the one hand, we consider a genuine hydrogen atom placed in vacuum near the WSM. In this case we find that when the topological term is switched off, the surface push out the atomic cloud (this is  understood from the electrostatic repulsion between the atomic electron and its image itself); however, in the presence of the topological term, the anomalous Hall effect contribution reverses this tendency, and the surface pulls in the atomic cloud (which we understand as a consequence of the force between the atomic current and the many Hall currents appearing in the bulk). {As expected, the focusing effect of the atomic cloud depends on the atom-surface distance: the cloud becomes a sharply focused peak as the atom approaches the interface. High resolution spectroscopy experiments, which has been successfully used to test the Casimir-Polder interaction between an atom in its ground state and metallic or dielectric samples (over 5 orders of magnitude for the potential strength) \cite{Oria_1991, PhysRevLett.68.3432, PhysRevLett.86.2766,PhysRevLett.83.5467}, can also be used to test the energy shift in the ground state for an atom near the WSM. Other nonspectroscopic methods, which has been used to test atom-surface interactions, could also be relevant in the present context, namely, atomic interferometry and quantum reflection. For example, a beam of atoms in the ground state traveling inside a cavity formed by material samples is sensitive to atom-surface interactions \cite{PhysRevLett.70.560}. The experiment measures the transmission (or rather the opacity) as a function of the separation between the plates: since for large separation one gets the geometrical expectation, by comparing with the result with a smaller separation one can therefore extract the information regarding the atom-surface interaction. On the other hand, quantum reflection has been used in Ref. \cite{PhysRevLett.86.987} to confirm experimentally the attractive character Casimir–van der Waals potential between an atom and metallic sample: for atoms approaching the sample at low incident kinetic energy, they are reflected well before reaching the interface, so the presence probability of the atoms remains vanishingly small around the minimum. Using a Weyl semimetallic target, this kind of experiment could be able to reveal the position of the minimum in the interaction potential, which strongly depends upon the topological contribution. } 

In order to enhance the effects of the topological terms upon a physical system, we also consider the case of an hydrogenic donor/acceptor impurity near the WSM, which as we know, within the effective-mass approximation, it is exactly equivalent to the quantum-mechanical hydrogen atom. The small mass (as compared with the electron mass) of a GaAs and GaP impurities make them an interesting possibility to test the anomalous Hall effect of the WSM in the present configuration. Indeed, we find that in the unperturbed case the atomic cloud is extended over a large region of space (a consequence of the small effective mass). However, when we turn on the topological coupling, the electron cloud is then strongly attracted towards the surface, thus exhibiting an interesting confinement effect. Due to the deformation of the atomic cloud, we expect also repercussions upon the atomic  polarization. Evaluating the corresponding expectation value we find that, in the unperturbed case, the polarization points away from the surface, while in the perturbed case the polarization flips and points towards the surface. The latter may have also deep implications in the electrical conductivity of doped semiconductors. This idea is being currently explored and will be reported elsewhere.


All in all, in this paper we have shown that the anomalous Hall effect, a distinctive manifestation of the topological charge of Weyl semimetals, induces significant effects upon the properties of an atom nearby, namely, energy shifts, the probability distribution and the polarization. This represents an alternative to the usual classical electrodynamics configurations in which the topological nontriviality manifests through optical observables. In the present case, being quantum observables, high-precision experiments increases the chance for detecting the topological features of the material.


\section*{Acknowledgements}
A.M.-R. has been partially supported by DGAPA-UNAM Project No. IA102722 and by Project CONACyT (M\'{e}xico) No. 428214 D.J.N. acknowledges A.V. Turbiner and J.C. Lopez-Vieyra for their advise in the design of the code architecture. J.D.C.-Y. thanks to Prof. Liliana Tirado and Prof. Gerardo Fonthal for their useful comments about the semiconductor's properties related to this work.

\appendix

\section{Boltzmann form of the conductivity} \label{AppKinetic}

In this Appendix, we derive an expression for the Drude part of the conductivity using kinetic theory.

The effective low-energy continuum Hamiltonian for time-reversal broken WSMs with a pair of nodes is given by
\begin{align}
H _{\chi} ({\bf{k}}) = b _{0 \chi} + \chi \hbar v _{F} \; {\boldsymbol{\sigma}} \cdot ({\bf{k}} - \chi {\bf{b}}) , \label{Hamiltonianwsm}
\end{align}
where $v _{F}$ is the Fermi velocity and $\chi = \pm 1$ is the chirality eigenvalue for a given Weyl node. The vector $\chi {\bf{b}}$ localizes the node with chirality $\chi$ from the origin at ${\bf{k}} = {\bf{0}}$. For simplicity, we choose coordinate axes such that ${\bf{b}}$ points along the Cartesian $z$-direction, i.e. ${\bf{b}} = b {\bf{e}} _{z}$. The energy dispersion is found to be
\begin{align}
E _{s \chi} ({\bf{k}}) = b _{0 \chi} + s \hbar v _{F} \sqrt{k _{x} ^{2} + k _{y} ^{2} + (k _{z} - \chi b) ^{2}} , \label{Energy}
\end{align}
where $s = \pm 1$ is the band index. 

In the semiclassical approach, the nonequilibrium quasiparticle current density driven by a single frequency electric field is given by
\begin{align}
    {\bf{J}} = - e \sum _{s = \pm 1} \sum _{\chi \pm 1} \int \frac{d^{3} {\bf{k}}}{(2 \pi ) ^{3}}  {\bf{v}} _{s \chi} \, f _{s \chi} ({\bf{k}},t) , \label{Current}
\end{align}
where ${\bf{v}} _{s \chi} = \tfrac{1}{\hbar} \nabla _{{\bf{k}}} E _{s \chi}$ is the band velocity and $f _{s \chi} ({\bf{k}},t)$ is the statistical distribution function of carriers in the phase space of position and crystal momentum. In the absence of an external field, the distribution is given by the Fermi-Dirac distribution $f _{s \chi} ^{\mbox{\scriptsize FD}}({\bf{k}}) = \left[ 1 +  \exp \left( \tfrac{E _{s \chi} - \mu _{\chi}}{k _{B} T} \right) \right] ^{-1}$, where $\mu _{\chi}$ is the chemical potencial for fermions with chirality $\chi$ and $k _{B}$ is the Boltzmann constant.  In the presence of external fields the nonequilibrium distribution solves the Boltzmann equation, which we solve up to linear order in the electric field and in the relaxation time approximation. Taking an input electric field of the form ${\bf{E}} = {\boldsymbol{\mathcal{E}}} e ^{\mathrm{i}  \omega t} + {\boldsymbol{\mathcal{E}}} ^{\ast} e ^{-\mathrm{i}  \omega t}$, the current (\ref{Current}) can be expressed as $J _{i} = \sigma _{ij} (\omega ) \mathcal{E} _{j} e ^{\mathrm{i}  \omega t} + \mbox{c.c.}$, where 
\begin{align}
\sigma _{ij} (\omega) = - \frac{e ^{2} \tau }{1 + \mathrm{i} \omega \tau} \sum _{s = \pm 1} \sum _{\chi = \pm 1} \int \frac{d^{3} {\bf{k}}}{(2 \pi ) ^{3}} v ^{i} _{s \chi} v ^{j} _{s \chi} \frac{\partial f _{s \chi} ^{\mbox{\scriptsize FD}} ({\bf{k}}) }{ \partial E _{s \chi}} , \label{LinearConductivity}
\end{align}
is the longitudinal conductivity tensor. For the sake of simplicity, we shall assume that the chemical potential $\mu _{\chi}$ is above the band-touching point of the node with chirality $\chi$, and hence we will concentrate on the transport phenomena of the conduction band (i.e. $s = +1$). Here, we will work at zero temperature $T=0$, such that $\partial f _{s \chi} ^{\mbox{\scriptsize FD}} ({\bf{k}}) / \partial E _{s \chi} = - \delta (\mu _{\chi} - E _{s \chi}) $. This means that the conductivity tensor (\ref{LinearConductivity}) is property of the Fermi surface. Defining the vector ${\bf{w}} _{\chi} = k _{x} \hat{\bf{e}} _{x} + k _{y} \hat{\bf{e}} _{y} + (k _{z} - \chi b ) \hat{\bf{e}} _{z}$, which is the crystalline momentum measured from the node with chirality $\chi$, the conductivity tensor (\ref{LinearConductivity}) simplifies to
\begin{align}
\sigma _{ij} (\omega) = \frac{e ^{2} \tau  v _{F} / \hbar }{1 + \mathrm{i} \omega \tau} \sum _{\chi = \pm 1} \frac{1}{\eta _{\chi} ^{2}} \int \frac{d^{3} {\bf{k}}}{(2 \pi ) ^{3}} w ^{i} _{\chi} w ^{j} _{ \chi} \delta (\eta _{\chi} - w ^{2} _{\chi}) , \label{LinearConductivity2}
\end{align}
where $\eta _{\chi} \equiv \tfrac{ \mu _{\chi} - b _{0 \chi} }{\hbar v _{F}}$. If we change the origin of the ${\bf{k}}$-space from $(0,0,0)$ to the node $(0,0,\chi b )$, the above integral simplifies to
\begin{align}
\sigma _{ij} (\omega) = \frac{e ^{2} \tau  v _{F} / \hbar}{1 + \mathrm{i} \omega \tau} \sum _{\chi = \pm 1} \frac{1}{\eta _{\chi} ^{2}} \int \frac{d^{3} {\bf{k}}}{(2 \pi ) ^{3}} k ^{i} k ^{j} \delta (\eta _{\chi} - k ^{2}) . \label{LinearConductivity3}
\end{align}
Evidently, the integral has spherical symmetry, and hence it can be written as
\begin{align}
\sigma _{ij} (\omega) = \frac{e ^{2} \tau v _{F} / \hbar}{1 + \mathrm{i} \omega \tau} \sum _{\chi = \pm 1} \frac{1}{\eta _{\chi} ^{2}} \frac{\delta _{ij}}{3} \int \frac{d^{3} {\bf{k}}}{(2 \pi ) ^{3}} k ^{2} \delta (\eta _{\chi} - k ^{2}) . \label{LinearConductivity4}
\end{align}
This integral is quite simple by using the properties of the Dirac delta function. The final result is then:
\begin{align}
\sigma _{ij} (\omega) = \frac{e ^{2} \tau}{1 + \mathrm{i} \omega \tau}  \frac{ \sum _{\chi = \pm 1}( \mu _{\chi} - b _{0 \chi}) ^{2}}{6 \pi ^{2} \hbar ^{3} v _{F}} \delta _{ij}  , \label{LinearConductivity5}
\end{align}
which is our expression for the conductivity tensor. In Eq.~(\ref{LongConductivity}) we present the limit $\omega \to 0 $ of the result (\ref{LinearConductivity5}).

{
\section{The existence of $z_c$}\label{Ap:zc}

As we stated in the main text, there is a value $z_c$ where the wave function cannot be normalized, i.e., there are not bound states. Such a behavior can be understood from the shape of the effective potential given by Eq.~(\ref{Eq:Veff_prolatas}). 

Let us concentrate in the ground state for which $m=0$, so that the condition for  turning points in the potential reads:
\bea
\nabla V_\text{eff}(\rho,z)={\bf{0}}.
\eea
From the analytical form of $V_\text{eff}(\rho,z)$ it is clear that the condition is satisfied if
\bea
\partial_zV_\text{eff}(\rho=0,z)=0.
\label{EcCritical}
\eea

\begin{figure}[h]
\centering
    \includegraphics[scale=0.55]{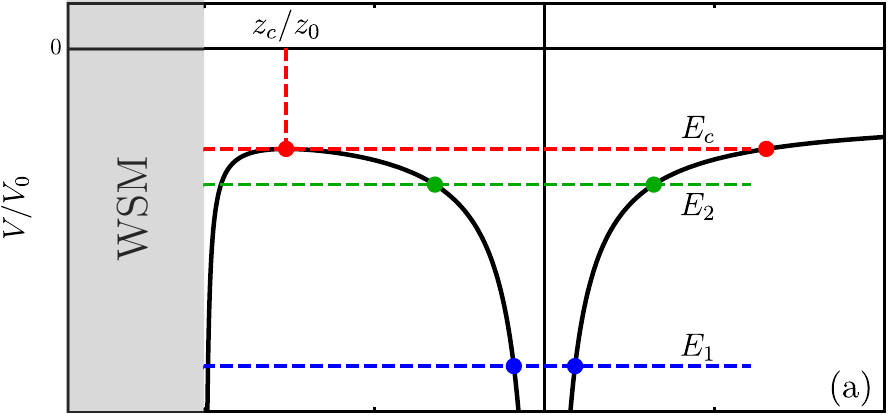}\\
    \,\includegraphics[scale=0.55]{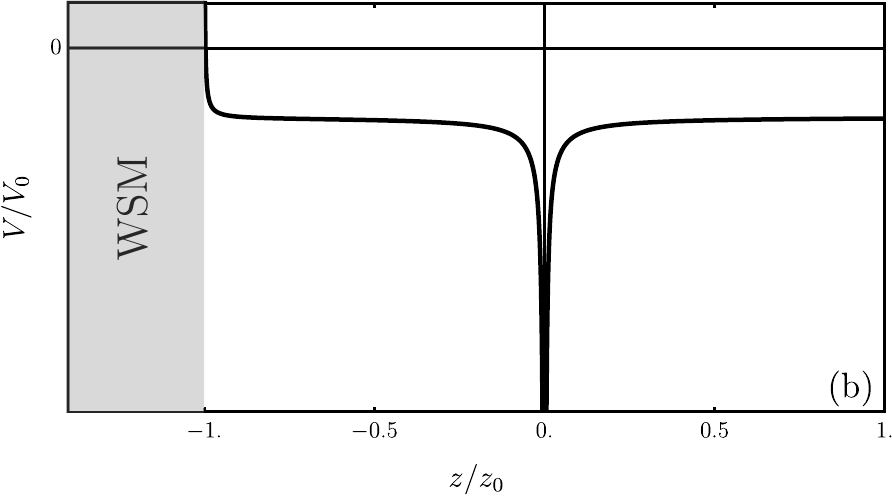}
    \caption{Scheme of the effective potential of Eq.~(\ref{Eq:Veff_prolatas}) with $m=0$ at $\rho=0$, as a function of $z$ for the two considered configurations of the relative permittivities: (a) $\epsilon_1>\epsilon_2$, and (b) $\epsilon_1<\epsilon_2$. In panel (a), $z_c$ indicates the critical point with its corresponding critical energy $E_c$. The energies $E_1$, and $E_2$ are arbitrary bound states. The dots are the turning points for each energy value. }
    \label{fig:potentialscketch}
\end{figure}

Figure~\ref{fig:potentialscketch} shows the dimensionless potential $V (\rho = 0, z) /V_{0}$ (where $V_{0} = e^{2}/z_{0}$) for $m=0$ as a function of the dimensionless distance $z/z_{0}$. At $z=0$, we observe the usual singular potential due to the nucleus-electron Coulomb interaction. Also, it decays properly for $z>0$. However, for $z<0$, the behavior of the effective potential is quite different due to the presence of the Weyl semimetal in the lower half-space. Moreover, close the surface $z=-z_{0}$, the effective potential is dominated by the Coulomb attraction/repulsion between the atomic electron and its image. For example, for $\mbox{sgn} (\epsilon_1-\epsilon_2) = 1$, the interaction is attractive (Fig.~\ref{fig:potentialscketch}-(a)), while $\mbox{sgn} (\epsilon_1-\epsilon_2) = -1$ implies a repulsive interaction (Fig.~\ref{fig:potentialscketch}-(b)). The latter exhibits the usual behavior of a Coulomb potential, so we concentrate on the former. Nevertheless, it is clear that the repulsive effect shown in Fig.~\ref{fig:potentialscketch}-(b) also has a critical distance so that close to the barrier imposed by the WSM, the atom gets ionized. 

In Fig.~\ref{fig:potentialscketch}-(a), it is clear the existence of a critical point $z_{c}$ in the region between the nucleus and the WSM surface, i.e., $z_{c}\in (-z_{0},0)$, which is determined from Eq.~(\ref{EcCritical}). Such a critical point defines a critical energy $E_c$. Therefore, for energies $E_k<E_c$ there exist bound quantum states, similar to the usual hydrogenic bound states. However, for $E_k>E_{c}$ bound states cannot be formed, and the wave functions given in Sec.~\ref{sec:Variational_Functions} ceases to be valid. An indicative of this is the non-normalizability of the wave function.

\begin{figure}[h!]
    \centering
    \includegraphics[scale=0.55]{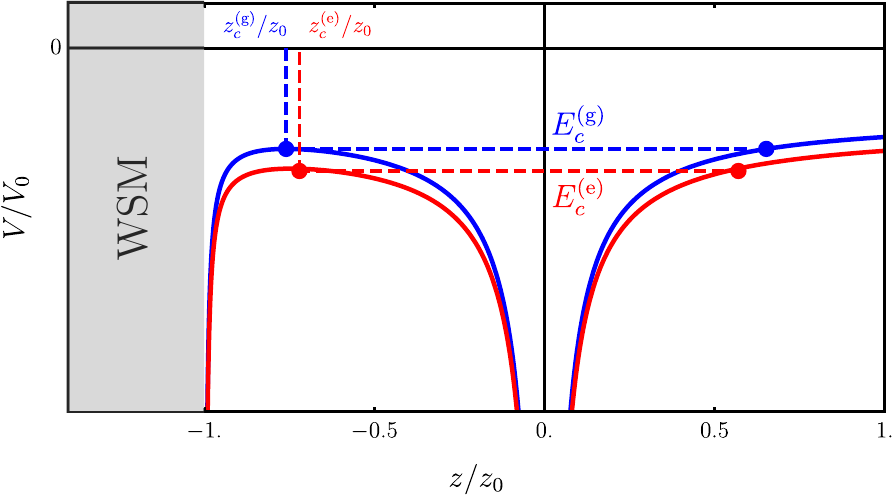}
    \caption{Critical points for the ground state $z_c^{\text{(g)}}$, and the first excited state $z_c^{\text{(e)}}$, each with its corresponding critical energy. }
    \label{fig:criticalpoint2}
\end{figure}

On the other hand, from the effective potential one can also understand the different behaviors in the plots of the probability density. Let us consider three different values of the energy, $E_{1}<E_{2}<E_{3}$, with $E_3\lesssim E_c$. All of them define two turning points, $z_{1}>0$ and $z_{2}<0$, so that the atomic electron bounces back and forth between these positions. For the energy $E _{1}$ one finds that $ z_{1} > \vert z_{2} \vert$, such that the electronic cloud is repelled by the WSM half-space. For the energy $E _{2}$ one gets $ z_{1} < \vert z_{2} \vert$ and hence the orientation is inverted, i.e. the atomic cloud is now attracted towards the WSM. For the energy $E_{3}$, which is very close to the critical energy $E_{c}$, the turning point $z_{2}$ approaches the critical point $z_{c}$. For an energy slightly above the critical energy, the electron ceases to be bounded and the atom becomes ionized.

The case of $m=1$ can be achieved in a similar fashion. The unique difference comes from the fact that the topological term modifies the critical point condition, i.e.,
\bea
0&=&\partial_zV_\text{eff}^{(\text{e})}(\rho=0,z)\nn\\
&=&\partial_zV^{(\text{e})}(\rho=0,z)\nn\\
&+&\frac{e ^{2} \epsilon _{1}  \hbar}{\mu c } \int _{0} ^{\infty} \!\! dk \frac{\alpha _{-} k ^{2} \, e ^{-k(z_{c} ^{(\text{e})}+2z _{0})}}{\epsilon _{1} \! \left( \alpha _{+} ^{2} + \alpha _{-} ^{2} \right) + \epsilon _{2} k ^{2} +  k  \alpha _{+} \! \left( \epsilon _{1} \! + \! \epsilon _{2} \right)},\nn\\
\eea
where (e) means ``excited'', and $\partial_zV^{(\text{e})}(\rho=0,z)$ is the effective potential for $m=0$. 

The latter equation shows explicitly that the ground state critical point $z_c^{(\text{g})}$ is different from the excited state one $z_c^{(\text{e})}$. Figure~\ref{fig:criticalpoint2} shows the effective potential for $m=0$ and $m=1$, with their corresponding critical points. As can be noticed, $|z_c^{(\text{g})}|>|z_c^{(\text{e})}|$ which means that the wave function normalization for $m=1$ is lost at larger distances from the WSM, compared with the case $m=1$. This is in concordance with Fig.~\ref{Fig:ExcitedStates} and can be generalized for other excited states.
}

\bibliography{Bib.bib}

\end{document}